\documentclass{article}
\usepackage{graphicx}
\usepackage{txfonts}
\usepackage{natbib}

\begin{document}
\sloppy

\setlength\textheight{10.35in}
\setlength\textwidth{7.2in}
\oddsidemargin -0.2in
\topmargin -1in

\def\aj{AJ}%
\def\actaa{Acta Astron.}%
\def\araa{ARA\&A}%
\def\apj{ApJ}%
\def\apjl{ApJ}%
\def\apjs{ApJS}%
\def\ao{Appl.~Opt.}%
\def\apss{Ap\&SS}%
\def\aap{A\&A}%
\def\aapr{A\&A~Rev.}%
\def\aaps{A\&AS}%
\def\azh{AZh}%
\def\baas{BAAS}%
\def\bac{Bull. astr. Inst. Czechosl.}%
\def\caa{Chinese Astron. Astrophys.}%
\def\cjaa{Chinese J. Astron. Astrophys.}%
\def\icarus{Icarus}%
\def\jcap{J. Cosmology Astropart. Phys.}%
\def\jrasc{JRASC}%
\def\mnras{MNRAS}%
\def\memras{MmRAS}%
\def\na{New A}%
\def\nar{New A Rev.}%
\def\pasa{PASA}%
\def\pra{Phys.~Rev.~A}%
\def\prb{Phys.~Rev.~B}%
\def\prc{Phys.~Rev.~C}%
\def\prd{Phys.~Rev.~D}%
\def\pre{Phys.~Rev.~E}%
\def\prl{Phys.~Rev.~Lett.}%
\def\pasp{PASP}%
\def\pasj{PASJ}%
\def\qjras{QJRAS}%
\def\rmxaa{Rev. Mexicana Astron. Astrofis.}%
\def\skytel{S\&T}%
\def\solphys{Sol.~Phys.}%
\def\sovast{Soviet~Ast.}%
\def\ssr{Space~Sci.~Rev.}%
\def\zap{ZAp}%
\def\nat{Nature}%
\def\iaucirc{IAU~Circ.}%
\def\aplett{Astrophys.~Lett.}%
\def\apspr{Astrophys.~Space~Phys.~Res.}%
\def\bain{Bull.~Astron.~Inst.~Netherlands}%
\def\fcp{Fund.~Cosmic~Phys.}%
\def\gca{Geochim.~Cosmochim.~Acta}%
\def\grl{Geophys.~Res.~Lett.}%
\def\jcp{J.~Chem.~Phys.}%
\def\jgr{J.~Geophys.~Res.}%
\def\jqsrt{J.~Quant.~Spec.~Radiat.~Transf.}%
\def\memsai{Mem.~Soc.~Astron.~Italiana}%
\def\nphysa{Nucl.~Phys.~A}%
\def\physrep{Phys.~Rep.}%
\def\physscr{Phys.~Scr}%
\def\planss{Planet.~Space~Sci.}%
\def\procspie{Proc.~SPIE}%
\let\astap=\aap
\let\apjlett=\apjl
\let\apjsupp=\apjs
\let\applopt=\ao
\def\aapr{Astr.~Astrophys.~Rev.}

\def\etal{{{\em et al.}\/}\ }
\def\Mpc{$h^{-1}$~{\rm  Mpc}}

\def\sm{\smallskip}

\input epsf.sty

\begin{twocolumn}


\title{\bf Dark Matter}


\bigskip


\author{ Jaan Einasto\\
Tartu Observatory, Estonia}



\maketitle

\tableofcontents

\section*{Summary}

We give a review of the development of the concept of dark matter.
The dark matter story passed through several stages on its way from a minor
observational puzzle to a major challenge for theory of elementary particles.  

We begin the review with the description of the discovery of the mass paradox
in our Galaxy and in clusters of galaxies. First hints of the problem appeared
already in 1930s and later more observational arguments were brought up, but
the issue of the mass paradox was mostly ignored by the astronomical community
as a whole.  In mid 1970s the amount of observational data was sufficient to
suggest the presence of a massive and invisible population around galaxies and
in clusters of galaxies.  The nature of the dark population was not clear at
that time, but the hypotheses of stellar as well as of gaseous nature of the
new population had serious difficulties.  These difficulties disappeared when
non-baryonic nature of dark matter was suggested in early 1980s.

The final break through came in recent years.  The systematic progress in the
studies of the structure of the galaxies, the studies of the large scale
structure  based on galaxy surveys, the analysis of the structure formation
after Big Bang, the chemical evolution of the Universe including the
primordial nucleosynthesis, as well as observations of the microwave
background showed practically beyond any doubt that the Universe actually
contains more dark matter than baryonic matter! In addition to the presence of
Dark Matter, recent observations suggest the presence of Dark Energy, which
together with Dark Matter and ordinary baryonic matter makes the total
matter/energy density of the Universe equal to the critical cosmological
density. Both Dark Matter and Dark Energy are the greatest challenges for
modern physics since their nature is unknown.

There are various hypotheses as for the nature of the dark matter particles,
and generally some form of weakly interactive massive particles (WIMPs) are
strongly favored. These particles would form a relatively cold medium thus
named Cold Dark Matter (CDM). The realization that we do not know the nature
of basic constituents of the Universe is a scientific revolution difficult to
comprehend, and the plan to hunt for the dark matter particles is one of the
most fascinating challenges for the future.

\section{ Dark Matter problem as a scientific revolution}

Almost all information on celestial bodies comes to us via photons. Most
objects are observed because they emit light. In other cases, like for example
in some nebulae, we notice dark regions against otherwise luminous background
which are due to absorption of light. Thus both light absorption and light
emission allow us to trace the matter in the Universe, and the study goes
nowadays well beyond the optical light.  Modern instruments have first
detected photon emission from astronomical bodies in the radio and infrared
regions of the spectrum, and later also in the X-ray and gamma-ray band, with
the use of detectors installed in space.

Presently available data indicate that astronomical bodies of different nature
emit (or absorb) photons in very different ways, and with very different
efficiency.  At the one end there are extremely luminous supernovae, when a
single star emits more energy than all other stars of the galaxy it belongs
to, taken together.  At the other extreme there are planetary bodies with a
very low light emission per mass unit.  The effectiveness of the emissivity
can be conveniently described by the mass-to-light ratio of the object,
usually expressed in Solar units in a fixed photometric system, say in blue
(B) light.  The examples above show that the mass-to-light ratio $M/L$ varies
in very broad range.  Thus a natural question arises: Do all astronomical
bodies emit or absorb light?  Observations carried out in the past century
have led us to the conclusion that the answer is probably NO.

Astronomers frequently determine the mass by studying the object emission.
However, the masses of astronomical bodies can be also determined directly,
using motions of other bodies (considered as test particles) around or within
the body under study.  In many cases such direct total mass estimates exceed
the estimated luminous masses of known astronomical bodies by a large
fraction. It is customary to call the hypothetical matter, responsible for
such mass discrepancy, {\bf Dark Matter}.

The realization that the presence of dark matter is a serious problem which
faces both modern astronomy and physics grew slowly but steadily. Early hints
did not call much attention.

{ The first indication for the possible presence of dark matter
  came from the dynamical study of our Galaxy. British astronomer
  James \citet{Jeans:1922fk} reanalyzed vertical motions of stars near
  the plane of the Galaxy, studied by the Dutch astronomer Jacobus
  \citet{Kapteyn:1922}.  Both astronomers calculated from these data
  the density of matter near the Sun. They also estimated the density
  due to all stars near the Galactic plane.  Kapteyn found that the
  spatial density of known stars is sufficient to explain the vertical
  motions. In contrast, Jeans results indicated the presence of two
  dark stars to each bright star. }

The second observation was made by Fritz \citet{Zwicky:1933}.  He
measured radial velocities of galaxies in the Coma cluster of
galaxies, and calculated the mean random velocities in respect to the
mean velocity of the cluster. Galaxies move in clusters along their
orbits; the orbital velocities are balanced by the total gravity of
the cluster, similar to the orbital velocities of planets moving
around the Sun in its gravitation field.  To his surprise Zwicky found
that orbital velocities are almost a factor of ten larger than
expected from the summed mass of all galaxies belonging to the
cluster.  Zwicky concluded that, in order to hold galaxies together in
the cluster, the cluster must contain huge amounts of some Dark
(invisible) matter.

The next hint of the dark matter existence came from cosmology. 

One of the cornerstones of the modern cosmology is the concept of an
expanding Universe.  From the expansion speed it is possible to
calculate the critical density of the Universe.  If the mean density
is less than the critical one, then the Universe has opened geometry;
if the mean density is larger than the critical, the Universe is
closed.  If the density has exactly the critical value, the spatial
geometry is flat. The mean density of the Universe can be estimated
using masses of galaxies and of the gas between galaxies. These
estimates show that the mean density of luminous matter (mostly stars
in galaxies and interstellar or intergalactic gas) is a few per cent
of the critical density. This estimate is consistent with the
constraints from the primordial nucleosynthesis of the light elements.

Another cornerstone of the classical cosmological model is the smooth
distribution of galaxies in space.  There exist clusters of galaxies,
but they contain only about one tenth of all galaxies.  Most of the
galaxies are more or less randomly distributed and are called field
galaxies.  This conclusion is based on counts of galaxies at various
magnitudes and on the distribution of galaxies in the sky.

Almost all astronomical data fitted well to these classical
cosmological paradigms until 1970s. Then two important analyses were
made which did not match the classical picture.  In mid 1970s first
redshift data covering all bright galaxies were available.  These data
demonstrated that galaxies are not distributed randomly as suggested
by earlier data, but form chains or filaments, and that the space
between filaments is practically devoid of galaxies.  Voids have
diameters up to several tens of megaparsecs.

At this time it was already clear that structures in the Universe form
by gravitational clustering, started from initially small fluctuations
of the density of matter. Matter ``falls'' to places where the density
is above the average, and ``flows away'' from regions where the
density is below the average. This gravitational clustering is a very
slow process.  In order to form presently observed structures, the
amplitude of density fluctuations must be at least one thousandth of
the density itself at the time of recombination, when the Universe
started to be transparent. The emission coming from this epoch was
first detected in 1965 as a uniform cosmic microwave background. When
finally the fluctuations of this background were measured by COBE
satellite they appeared to be two orders of magnitude lower than
expected from the density evolution of the luminous mass.
 
The solution of the problem was suggested independently by several
theorists.  In early 1980s the presence of dark matter was confirmed
by many independent sources: the dynamics of the galaxies and stars in
the galaxies, the mass determinations based on gravitational lensing,
and X-ray studies of clusters of galaxies.  If we suppose that the
dominating population of the Universe -- Dark Matter -- is not made of
ordinary matter but of some sort of non-baryonic matter, then density
fluctuations can start to grow much earlier, and have at the time of
recombination the amplitudes needed to form structures.  The
interaction of non-baryonic matter with radiation is much weaker than
that of ordinary matter, and radiation pressure does not slow the
early growth of fluctuations.

The first suggestions for the non-baryonic matter were particles well
known at that time to physicists -- neutrinos. However, this scenario
soon led to major problems. Neutrinos move with very high velocities
which prevents the formation of small structures as galaxies.  Thus
some other hypothetical non-baryonic particles were suggested, such as
axions.  The essential property of these particles is that they have
much lower velocities. Because of this the new version of Dark Matter
was called Cold, in contrast to neutrino-dominated Hot Dark Matter.
Numerical simulations of the evolution of the structure of the
Universe confirmed the formation of filamentary superclusters and
voids in the Cold Dark Matter dominated Universe.

The suggestion of the Cold Dark Matter has solved most problems of the
new cosmological paradigm.  The actual nature of the CDM particles is
still unknown.  Physicists have attempted to discover particles which
have properties needed to explain the structure of the Universe, but
so far without success.

One unsolved problem remained.  Estimates of the matter density
(ordinary $+$ dark matter) yield values of about 0.3 of the critical
density.  This value -- not far from unity but definitely smaller than
unity -- is neither favored by theorists nor by the data, including
the measurements of the microwave background, the galaxy dynamics and
the expansion rate of the Universe obtained from the study of
supernovae.  To fill the matter/energy density gap between unity and
the observed matter density it was assumed that some sort of vacuum
energy exists.  This assumption is not new: already Einstein added to
his cosmological equations a term called the Lambda-term.  About ten
years ago first direct evidence was found for the existence of the
vacuum energy, presently called Dark Energy.  This discovery has
filled the last gap in the modern cosmological paradigm.

In the International Astronomical Union (IAU) symposium on Dark Matter
in 1985 in Princeton, \citet{Tremaine:1987} characterized the
discovery of the dark matter as a typical scientific revolution,
connected with changes of paradigms.  \citet{Kuhn:1970} in his book
{\em The Structure of Scientific Revolutions} discussed in detail the
character of scientific revolutions and paradigm changes. There are
not so many areas in modern astronomy where the development of ideas
can be described in these terms, thus we shall discuss the Dark Matter
problem also from this point of view. Excellent reviews on the dark
matter and related problems are given by \citet{Faber:1979},
\citet{Trimble:1987}, \citet{Srednicki:1990}, \citet{Turner:1991},
\citet{Silk:1992}, \citet{van-den-Bergh:2001}, \citet{Ostriker:2003a},
\citet{Rees:2003}, \citet{Turner:2003}, 
\citet{Trimble:2010n} and \citet{Sanders:2010}, see
also proceedings by \citet{Longair:1978}, and \citet{Kormendy:1987}.

\section{Early evidence of the existence of dark matter}

\subsection{Local Dark Matter}

The dynamical density of matter in the Solar vicinity can be estimated
using vertical oscillations of stars around the galactic plane.  The
orbital motions of stars around the galactic center play a much
smaller role in determining the local density.  Ernst
\citet{Opik:1915} found that the summed contribution of all known
stellar populations (and interstellar gas) is sufficient to explain
the vertical oscillations of stars -- in other words, there is no need
to assume the existence of a dark population.  { Similar analyses
  were made by \citet{Kapteyn:1922} and \citet{Jeans:1922fk}, who used
  the term ``Dark Matter'' to denote the invisible matter which existence
  is suggested by its gravity only.  Kapteyn found for the dynamical
  density of matter near the Sun 0.099~$M_\odot/pc^3$,
  Jeans got 0.143 in the same units.

  The next very careful determination of the matter density near the
  Sun was made by Jan \citet{Oort:1932}. His analysis indicated that
  the total density, found from dynamical data, is
  0.092~$M_\odot/pc^3$, and the density of stars, including expected
  number of white dwarfs, is approximately equal to the dynamical
  density. He concluded that the total mass of nebulous or meteoric
  dark matter near the Sun is very small.

The local density of matter has been re-determined by various authors
many times.  Grigori \citet{Kuzmin:1952, Kuzmin:1955} and his students
Heino \citet{Eelsalu:1959} and Mihkel \citet{Joeveer:1972,
  Joeveer:1974} confirmed the earlier results by \"Opik, Kapteyn and
Oort. A number of other astronomers, including more recently
\citet{Oort:1960}, John \citet{Bahcall:1980, Bahcall:1984,
  Bahcall:1987}, found results in agreement with the Jeans result.
Their results mean that the amount of invisible matter in the Solar
vicinity should be approximately equal to a half of the amount of
visible matter. This discussion was open until recently; we will
describe the present conclusions below. }

For long time no distinction between local and global dark matter was
made. The realization, that these two types of dark matter have very
different properties and nature came from the detailed study of
galactic models, as we shall discuss below \citep{Einasto:1974a}.

\subsection{Global Dark Matter  -- clusters,  groups and galaxies}

A different mass discrepancy was found by Fritz
\citet{Zwicky:1933}. He measured redshifts of galaxies in the Coma
cluster and found that the velocities of individual galaxies with
respect to the cluster mean velocity are much larger than those
expected from the estimated total mass of the cluster, calculated from
masses of individual galaxies.  The only way to hold the cluster from
rapid expansion is to assume that the cluster contains huge quantities
of some invisible dark matter. According to his estimate the amount of
dark matter in this cluster exceeds the total mass of cluster galaxies
at least tenfold, probably even more. 

{ \citet{Smith:1936mn} measured radial velocities of 30 galaxies in
  the Virgo cluster and confirmed the Zwicky result that the total
  dynamical mass of this cluster exceeds considerably the estimated
  total mass of galaxies.  This conclusion was again confirmad by
  \citet{Zwicky:1937uz}, who discussed masses of galaxies and clusters
  in detail.  As characteristic in scientific revolutions, early
  indications of problems in current paradigms are ignored by the
  community, this happened also with the Zwicky's discovery.  

  A certain discrepancy was detected between masses of individual
  galaxies and masses of pairs and groups of galaxies
  (\citet{Holmberg:1937uq}, \citet{Page:1952, Page:1959, Page:1960}).
  The conventional approach for the mass determination of pairs and
  groups of galaxies is statistical.  The method is based on the
  virial theorem and is almost identical to the procedure used to
  calculate masses of clusters of galaxies.  Instead of a single pair
  or group often a synthetic group is used consisting of a number of
  individual pairs or groups.  These determinations yield for the
  mass-to-light ratio (in blue light) the values $M/L_B = 1 \dots 20$
  for spiral galaxy dominated pairs, and $M/L_B = 5 \dots 90$ for
  elliptical galaxy dominated pairs (for a review see
  \citet{Faber:1979}).  These ratios are larger than found from local
  mass indicators of galaxies (velocity dispersions at the center and
  rotation curves of spiral galaxies).  However, it was not clear how
  serious is the discrepancy between the masses found using global or
  local mass indicators.}

A completely new approach in the study of masses of systems of
galaxies was applied by \citet{Kahn:1959}.  They paid attention to the
fact that most galaxies have positive redshifts as a result of the
expansion of the Universe; only the Andromeda galaxy (M31) has a
negative redshift of about 120 km/s, directed toward our Galaxy. This
fact can be explained, if both galaxies, M31 and our Galaxy, form a
physical system. A negative radial velocity indicates that these
galaxies have already passed the apogalacticon of their relative orbit
and are presently approaching each other. From the approaching
velocity, the mutual distance, and the time since passing the
perigalacticon (taken equal to the present age of the Universe), the
authors calculated the total mass of the double system. They found
that $M_{tot} \geq 1.8 \times 10^{12}~ M_{\odot}$.  The conventional
masses of the Galaxy and M31 were estimated to be of the order of $2
\times 10^{11}~ M_{\odot}$.  In other words, the authors found
evidence for the presence of additional mass in the Local Group of
galaxies.  The authors suggested that the extra mass is probably in
the form of hot gas of temperature about $5 \times 10^5$ K.  Using
more modern data \citet{Einasto:1982} made a new estimate of the total
mass of the Local Group, using the same method, and found the total
mass of $4.5 \pm 0.5 \times 10^{12} M_{\odot}$. This estimate is in
good agreement with recent determinations of the sum of masses of M31
and the Galaxy including their dark halos (see below).

{ In 1961 during the International Astronomical Union (IAU) General
  Assembly a symposium on problems in extragalactic research was
  organised \citep{McVittie:1962pd}, and a special meeting to discuss
  the stability of clusters of galaxies \citep{Neyman:1961zr}. The last
  meeting was the first wide discussion of the mass discrepancy in
  clusters of galaxies.}  Here the hypothesis of Victor
\citet{Ambartsumian:1961} on the stability and possible expansion of
clusters was analysed in detail.  Sidney \citet{van-den-Bergh:1961a,
  van-den-Bergh:1962a} drew attention to the fact that the dominating
population in elliptical galaxies is the bulge consisting of old
stars, indicating that cluster galaxies are old.  It is very difficult
to imagine how old cluster galaxies could form an instable and
expanding system.  These remarks did not find attention and the
problem of the age and stability of clusters remained open. The
background of this meeting and views of astronomers supporting these
and some alternative solutions were described by
\citet{Trimble:1995c}, \citet{van-den-Bergh:2001},
\citet{Sanders:2010} and \citet{Trimble:2010n}.

\begin{figure*}[ht]
\centering
\resizebox{0.8\textwidth}{!}
{\includegraphics{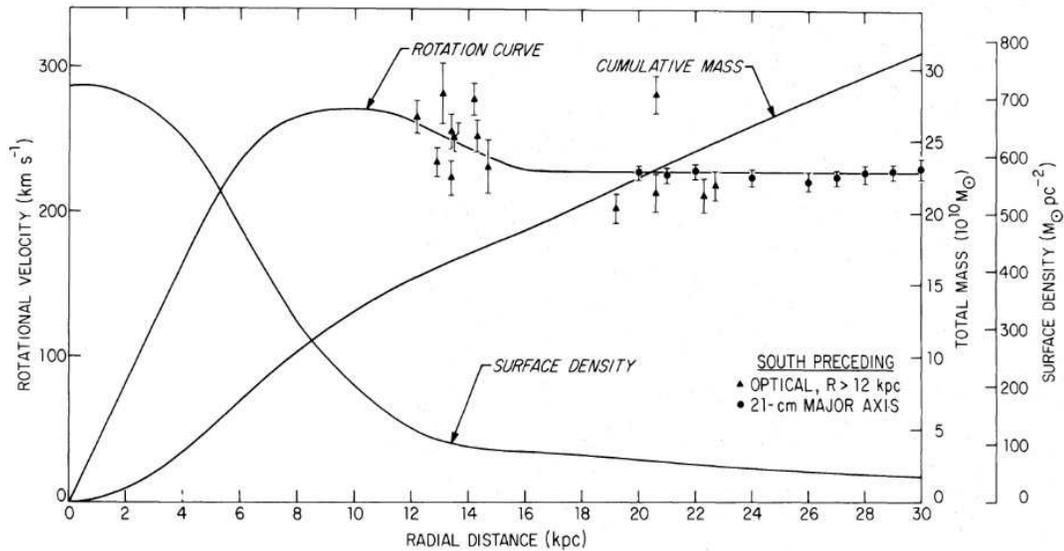}}
\caption{The rotation curve of M31 by \citet{Roberts:1975a}.  The filled
  triangles show the optical data from \citet{Rubin:1970}, the
  filled circles show the 21-cm measurements made with the 300-ft radio
  telescope  (reproduced by permission of the AAS and the author).  }
\label{fig:M31}
\end{figure*}

\subsection{Rotation curves of galaxies}

Another problem with the distribution of mass and mass-to-light ratio
was detected in spiral galaxies.  \citet{Babcock:1939} obtained
spectra of the Andromeda galaxy M31, and found that in the outer
regions the galaxy is rotating with an unexpectedly high velocity, far
above the expected Keplerian velocity.  He interpreted this result
either as a high mass-to-light ratio in the periphery or as a strong
dust absorption.  \citet{Oort:1940} studied the rotation and surface
brightness of the edge-on S0 galaxy NGC 3115, and found in the outer
regions a mass-to-light ratio $\sim 250$.  

{ After World War II there were numerous German radar dishes in the
  Dutch territory, and Oort and his colleagues understood that these
  devices can be used to detect radio waves from astronomical
  objects. His student van de Hulst has calculated that hydrogen
  emits radio waves in 21-cm, and this emission can be used to detect
  interstellar hydrogen and to measure its velocity. The first goal
  was to measure the radio emission from our own Galaxy
\citep{van-de-Hulst:1954vg}.  The next
  goal was the Andromeda galaxy M31. \citet{van-de-Hulst:1957um} found
  that the neutral hydrogen emitting the 21-cm line extends much
  farther than the optical image. They were able to measure the
  rotation curve of M31 up to about 30 kpc from the center, confirming the
  global value of $M/L \sim 20$ versus $M/L \sim 2$ in the central
  region.  

On the other hand, \citet{Schwarzschild:1954ys} analysed all available
data on mass-to-luminosity ratio in galaxies, and found that within
the optically visible disk $M/L$ is approximately constant, i.e. mass
follows light. In elliptical galaxies this ratio is higher than in
spiral galaxies. 

About ten years later Morton \citet{Roberts:1966dt} made a
  new 21-cm hydrogen line survey of M31 using the National Radio
  Astronomy Observatory large 300-foot telescope. The flat rotation
  curve at large radii was confirmed with much higher accuracy. He
  constructed also a mass distribution model of M31.

  Astronomers having access to large optical telescopes continued to
  collect dynamical data on galaxies. The most extensive series of
  optical rotation curves of galaxies was made by Margaret and
  Geoffrey Burbidge, starting from
  \citet{Burbidge:1959fx,Burbidge:1959hs}, and including normal and
  barred spirals as well as some ellipticals.  For all galaxies
  authors calculated mass distribution models, for spiral galaxies
  rotation velocities were approximated by a polynom.  They found that
  in most galaxies within visible images the mean $M/L \sim 3$.  

Subsequently, \citet{Rubin:1970} and \citet{Roberts:1973fb} derived the
rotation curve of M31 up to a distance $\sim 30$ kpc, using optical
and radio data, respectively. } The rotation speed rises slowly with
increasing distance from the center of the galaxy and remains almost
constant over radial distances of 16--30 kpc, see Fig.~\ref{fig:M31}.

The rotation data allow us to determine the distribution of mass, and
the photometric data -- the distribution of light. Comparing both
distributions one can calculate the local value of the mass-to-light
ratio.  In the periphery of M31 and other galaxies studied the local
value of $M/L$, calculated from the rotation and photometric data,
increases very rapidly outwards, if the mass distribution is
calculated directly from the rotation velocity.  In the periphery old
metal-poor halo-type stellar populations dominate.  These metal-poor
populations have a low $M/L \approx 1$ (this value can be checked
directly in globular clusters which contain similar old metal-poor
stars as the halo).  In the peripheral region the luminosity of a
galaxy drops rather rapidly, thus the expected circular velocity
should decrease according to the Keplerian law.  In contrast, in the
periphery the rotation speeds of galaxies are almost constant, which
leads to very high local values of $M/L > 200$ near the last points
with a measured rotational velocity.

Two possibilities were suggested to solve this controversy.  One
possibility is to identify the observed rotation velocity with the
circular velocity. But in this case an explanation for a very high
local $M/L$ should be found.  To explain this phenomenon it was
suggested that in outer regions of galaxies low-mass dwarf stars
dominate \citep{Oort:1940, Roberts:1975}.  The other possibility is to
assume that in the periphery of galaxies there exist non-circular
motions which distort the rotation velocity.

To make a choice between the two possibilities for solving the mass
discrepancy in galaxies more detailed models of galaxies were needed.
In particular, it was necessary to take into account the presence of
galactic stellar populations with different physical properties (age,
metal content, colour, $M/L$ value, spatial and kinematical
structure).

\subsection{Mass paradox in galaxies from Galactic models}

Classical models of elliptical galaxies were found from luminosity
profiles and calibrated using either central velocity dispersions, or
motions of companion galaxies.  The luminosity profiles of disks were
often approximated by an exponential law, and bulge and halo dominated
ellipticals by the \citet{de-Vaucouleurs:1953a} law.

Models of spiral galaxies were constructed using rotation velocities.
As a rule, the rotation velocity was approximated by some simple
formula, such as the \citet{Bottlinger:1933} law 
\citep{Roberts:1966dt}, or a polynomial  \citep{Burbidge:1959hs}.
The other possibility was to approximate the spatial density
(calculated from the rotation data) by a sum of ellipsoids of constant
density (the \citet{Schmidt:1956} model).  In the first case there
exists a danger that, if the velocity law is not chosen well, then the
density in the periphery of the galaxy may have unrealistic values
(negative density or too high density, leading to an infinite total
mass).  If the model is built by superposition of ellipsoids of
constant density, then the density is not a smooth function of the
distance from the center of the galaxy.  To avoid these difficulties
\citet{Kuzmin:1952a, Kuzmin:1956} developed models with a continuous
change of the spatial density, and applied the new technique to M31
and our Galaxy.  His method allows us to apply this approach also for
galaxies consisting of several populations.

A natural generalization of classical galactic models is the use of
all available observational data for spiral and elliptical galaxies,
both photometric data on the distribution of color and light, and
kinematical data on the rotation and/or velocity dispersion.  Further,
it is natural to apply identical methods for modeling of galaxies of
different morphological type (including our own Galaxy), and to
describe explicitly all major stellar populations, such as the bulge,
the disk, the halo, as well as the flat population in spiral galaxies,
consisting of young stars and interstellar gas.

All principal descriptive functions of galaxies (circular velocity,
gravitational potential, projected density) are simple integrals of
the spatial density. Therefore it is natural to apply for the spatial
density $\rho(a)$ of galactic populations a simple generalized
exponential expression \citep{Einasto:1965}:
\begin{equation}
\rho(a) = \rho(0) \exp\left(-(a/a_{0})^{1/N}\right),
\label{explaw}
\end{equation}
where $a$ is the semi-major axis of the isodensity ellipsoid, $a_0$ is
the effective radius of the population, and $N$ is a structural
parameter, determining the shape of the density profile. This
expression (called the Einasto profile) can be used for all galactic
populations, including dark halos. The case $N=4$ corresponds to the
de Vaucouleurs density law for spheroidal populations, $N=1$
corresponds to the exponential density law for disk.  { A similar
  profile has been used by \citet{Sersic:1968dz} for the projected
  density of galaxies and their populations; in this case the
  parameter $N$ is called the Sersic index.}

Multi-component models for spiral and elliptical galaxies using
photometric data were constructed by \citet{Freeman:1970}.  To combine
photometric and kinematic data, mass-to-light ratios of galactic
populations are needed. Luminosities and colors of galaxies in various
photometric systems result from the physical evolution of stellar
populations that can be modeled. { The study of the chemical
  evolution of galaxies was pioneered by \citet{Schwarzschild:1953qh}
  and \citet{Cameron:1971}.  Detailed models of the physical and
  chemical evolution of galaxies were constructed by
  \citet{Tinsley:1968}.

  Combined population and physical evolution models were calculated
  for a representative sample of galaxies by \citet{Einasto:1972}. It
  is natural to expect, that in similar physical conditions the
  mass-to-luminosity ratio $M_i/L_i$ of the population $i$ has similar
  values in different stellar systems (star clusters, galactic
  populations). Thus we can use compact stellar populations (star
  clusters and central cores of galaxies), to estimate $M_i/L_i$
  values for the main galactic populations.
  
Results of these calculations were reported at the First European
Astronomy Meeting in Athens in September 1972 by
\citet{Einasto:1974a}.  The main conclusion was: it is impossible to
reproduce the rotation data by known stellar populations only.  The
only way to eliminate the conflict between photometric and rotational
data was {\em to assume the presence of an unknown almost spherical
  population with a very high value of the mass-to-light ratio, large
  radius and mass}.  To avoid confusion with the conventional stellar
halo, the term ``corona'' was suggested for the massive population.
Thus, the detailed modeling confirmed earlier results obtained by
simpler models.  But here we have one serious difficulty -- no known
stellar population has so large a $M/L$ value.}

Additional arguments for the presence of a spherical massive
population in spiral galaxies came from the stability criteria against
bar formation, suggested by \citet{Ostriker:1973}. Their numerical
calculations demonstrated that initially very flat systems become
rapidly thicker (during one revolution of the system) and evolve to a
bar-like body. In real spiral galaxies a thin population exists, and
it has no bar-like form.  In their concluding remarks the authors
write: {\em ``Presumably even Sc and other relatively 'pure' spirals
  must have some means of remaining stable, and the possibility exists
  that those systems also have very large, low-luminosity halos.  The
  picture developed here agrees very well with the fact, noted by
  several authors (see, for example, \citet{Rogstad:1972}), that the
  mass-to-light ratio increases rapidly with distance from the center
  in these systems; the increase may be due to the growing dominance
  of the high mass-to-light halo over the low mass-to-light ratio
  disk. It also suggests that the total mass of such systems has been
  severely underestimated.  In particular, the finding of
  \citet{Roberts:1973fb} that the rotation curves of several nearby
  spirals become flat at large distances from the nucleus may indicate
  the presence of very extended halos having masses that diverge
  rapidly [M(r) prop to r] with distance.''}

\section{Dark Matter in astronomical data}

Modern astronomical methods yield a variety of independent information
on the presence and distribution of dark matter.  For our Galaxy, the
basic data are the stellar motions perpendicular to the plane of the
Galaxy (for the local dark matter), the motions of star and gas
streams and the rotation (for the global dark matter).  Important
additional data come from gravitational microlensing by invisible
stars or planets. In nearby dwarf galaxies the basic information comes
from stellar motions.  In more distant and giant galaxies the basic
information comes from the rotation curves and the X-ray emission of
the hot gas surrounding galaxies.  In clusters and groups of galaxies
the gravitation field can be determined from relative motions of
galaxies, the X-ray emission of hot gas and gravitational lensing.
Finally, measurements of fluctuations of the Cosmic Microwave
Background (CMB) radiation in combination with data from type Ia
supernovae in nearby and very distant galaxies yield information on
the curvature of the Universe that depends on the amount of Dark
Matter and Dark Energy.

Now we shall discuss these data in more detail.

\subsection{Stellar motions}

The local mass density near the Sun can be derived from vertical
oscillations of stars near the galactic plane, as was discussed
before.  Modern data by \citet{Kuijken:1989b, Gilmore:1989} have
confirmed the results by Kuzmin and his collaborators.  Thus we come
to the conclusion that {\em there is no evidence for the presence of
  large amounts of dark matter in the disk of the Galaxy}.  If there
is some invisible matter near the galactic plane, then its amount is
small, of the order of 15 percent of the total mass density.  The
local dark matter is probably baryonic (low--mass stars or jupiters),
since non-baryonic matter is dissipationless and cannot form a highly
flattened population.  Spherical distribution of the local dark matter
{ (in quantities suggested by \citet{Oort:1960} and
\citet{Bahcall:1987})} is excluded since in this case the total mass of
the dark population would be very large and would influence also the
rotational velocity of the Galaxy at the location of the Solar System.

Additional information of the distribution of mass in the outer part
of the Galaxy comes from streams of stars and gas.  One of the streams
discovered near the Galaxy is the Magellanic Stream of gas which forms
a huge strip and connects the Large Magellanic Cloud (LMC) with the
Galaxy \citep{Mathewson:1974}.  Model calculations emphasize that this
stream is due to an encounter of the LMC with the Galaxy.  Kinematical
data for the stream are available and support the hypothesis on the
presence of a massive halo surrounding the Galaxy
\citep{Einasto:1976a}.  Recently, streams of stars have been
discovered within the Galaxy as well as around our giant neighbor M31.
Presently there are still few data on the kinematics of these streams.

Several measurements of the dark mass halo were also performed using
the motion of the satellite galaxies or the globular
clusters. Measurements indicate the mass of the dark halo of about $2
\times 10^{12} M_{\odot}$.

However, significant progress is expected in the near future. The
astronomical satellite GAIA (to fly in 2011) is expected to measure
distances and photometric data for millions of stars in the Galaxy.
When these data are available, more information on the gravitation
field of the Galaxy can be found.

The motion of individual stars or gaseous clouds can be also studied
in nearby dwarf galaxies. Determination of the dark halo was performed
for over a dozen of them.  Some of the newly discovered dwarfs, coming
from the Sloan Digital Sky Survey, are very under-luminous but equally
massive as the previously known dwarf galaxies in the Milky Way
vicinity, which makes them good candidates for extreme examples of
dark matter dominated objects. Also the studies of the disruption rate
of these galaxies due to the interaction with the Milky Way impose
limits to the amount of dark mass in these objects.  The results
indicate that the dark matter in these systems exceeds by a factor a
few the mass of stars.

\subsection{Dynamics and morphology of companion galaxies}

The rotation data available in early 1970s allowed the determination
the mass distribution in galaxies up to their visible edges.  In order
to find how large and massive galactic coronas or halos are, more
distant test particles are needed.  If halos are large enough, then in
pairs of galaxies the companion galaxies are located inside the halo,
and their relative velocities can be used instead of the galaxy
rotation velocities to find the distribution of mass around giant
galaxies.  { This test was made by \citet{Einasto:1974} (see
  Fig.~\ref{fig:MRpair} ), and reported in a cosmology winter-school
  near the Elbrus mountain on January 1974, organized by Yakov
  Zeldovich. Zeldovich and his group had been working over 15 years to
  find basic physical processes for the formation and evolution of the
  structure of the Universe. For them the presence of a completely new massive
  non-stellar population was a great surprise, and caused an avalanche
  of new studies to find its properties and physical nature
  (\citet{Ozernoi:1974sr},  \citet{Jaaniste:1975}, 
  \citet{Komberg:1975},  \citet{Bobrova:1975pt}, \citet{Zeldovich:1975}
  among others).  }

\begin{figure}[ht]
\centering
\vspace{3mm}
\resizebox{0.42\textwidth}{!}
{\includegraphics{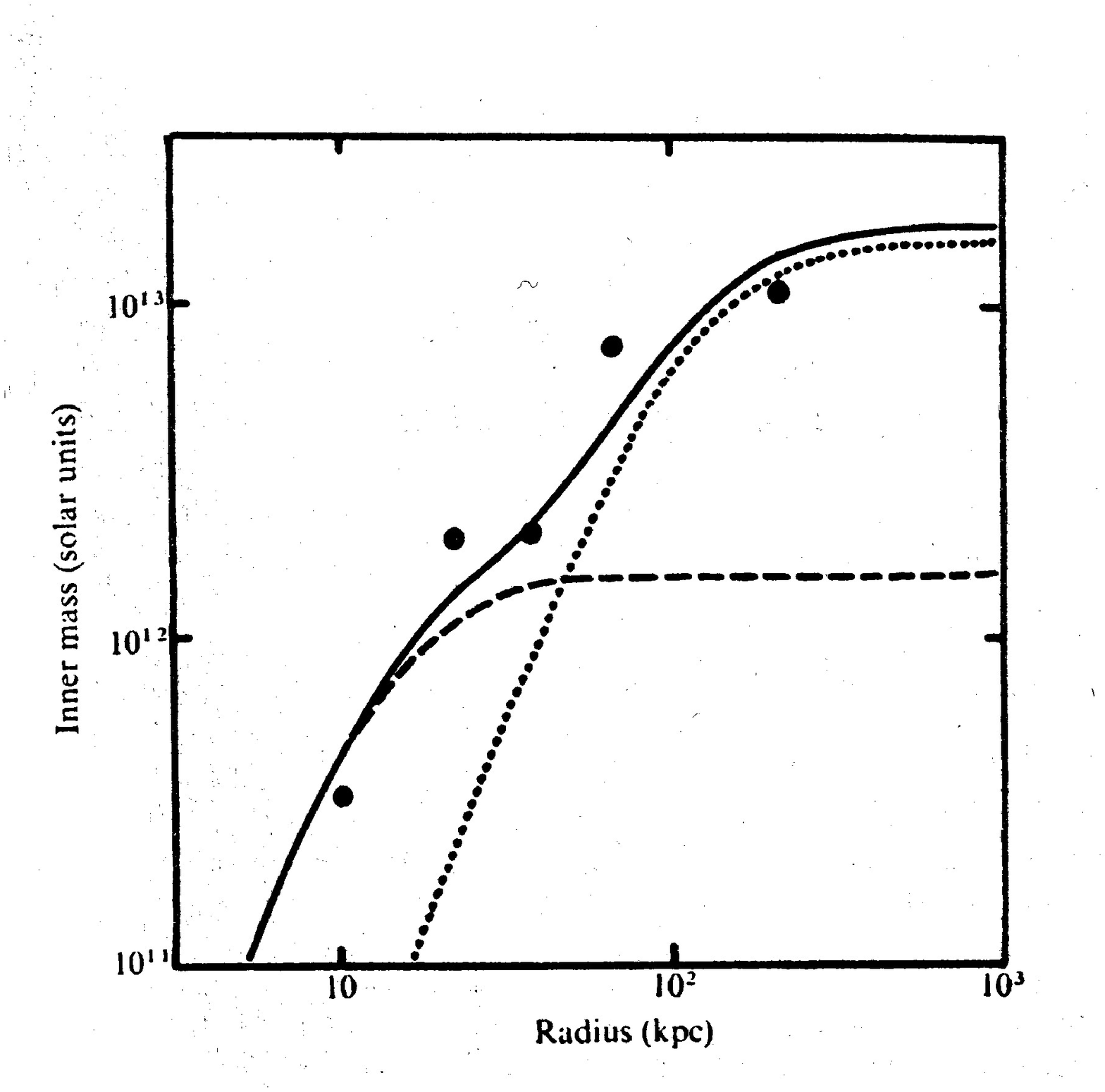}}
\caption{The mean internal mass $M(R)$ as a function of the radius $R$ from the
  main galaxy in 105 pairs of galaxies (dots).  Dashed line shows the
  contribution of visible populations, dotted line the contribution of the
  dark corona, solid line the total distribution \citep{Einasto:1974}.
}
\label{fig:MRpair}
\end{figure}

A similar study was made independently by \citet{Ostriker:1974}, see
Fig.~\ref{fig:OPY74}.  The paper by Ostriker {\em et al.} begins with
the statement: {\em ``There are reasons, increasing in number and
  quality, to believe that the masses of ordinary galaxies may have
  been underestimated by a factor of 10 or more''}. The closing
statement of the Einasto {\em et al.} paper is: {\em ``The mass of
  galactic coronas exceeds the mass of populations of known stars by
  one order of magnitude. According to new estimates the total mass
  density of matter in galaxies is 20\% of the critical cosmological
  density.''}  The bottom line in both papers was: since the data
suggest that all giant galaxies have massive halos/coronas, dark
matter must be the dynamically dominating population in the whole
Universe.

Results of these papers were questioned by \citet{Burbidge:1975}, who
noticed that satellites may be optical.  To clarify if the companions
are true members of the satellite systems, \citet{Einasto:1974b}
studied the morphology of companions.  They found that companion
galaxies are segregated morphologically: elliptical (non--gaseous)
companions lie close to the primary (host) galaxy whereas spiral and
irregular (gaseous) companions of the same luminosity have larger
distances from the primary galaxy.  The elliptical/spiral segregation
line from the primary galaxy depends on the luminosity of the
satellite galaxy, see Fig.~\ref{fig:segr}.  This result shows, first of
all, that the companions are real members of these systems -- random
by-fliers cannot have such properties.  Second, this result
demonstrates that diffuse matter has an important role in the
evolution of galaxy systems.  Morphological properties of companion
galaxies can be explained, if we assume that (at least part of) the
corona is gaseous.

\begin{figure}[ht]
\centering
\resizebox{0.45\textwidth}{!}
{\includegraphics{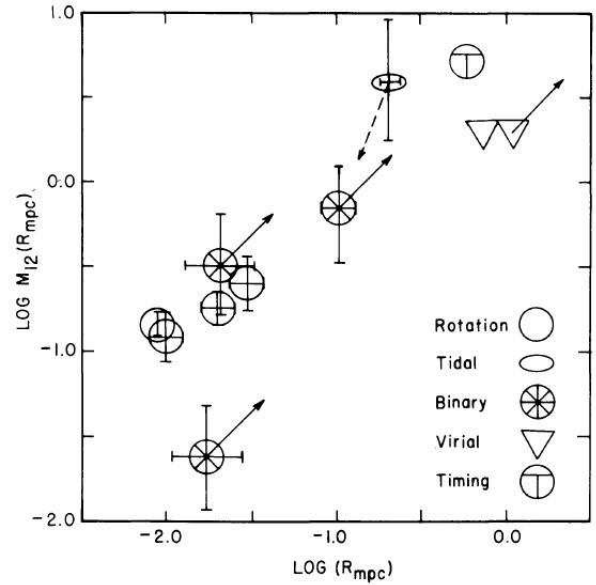}}
\caption{Masses (in units $10^{12}~M_\odot$) of local giant galaxies
  \citep{Ostriker:1974} (reproduced by permission of the AAS and authors). 
}
\label{fig:OPY74}
\end{figure}

Additional arguments in favor of physical connection of companions
with their primary galaxies came from the dynamics of small
groups. Their mass distribution depends on the morphology: in systems
with a bright primary galaxy the density (found from kinematical data)
is systematically higher, and in elliptical galaxy dominated systems
it is also higher.  The mass distribution found from the kinematics of
group members smoothly continues the mass distribution of the primary
galaxies, found from rotation data \citep{Einasto:1976}.

\subsection{Extended rotation curves of galaxies}

The dark matter problem was discussed in 1975 at two conferences, in
January in Tallinn \citep{Doroshkevich:1975} and in July in Tbilisi.
The central problems discussed in Tallinn were: Deuterium abundance
and the mean density of the universe \citep{Zeldovich:1975}, What is
the physical nature of the dark matter?  and: What is its role in the
evolution of the Universe?  Two basic models were suggested for
coronas: faint stars or hot gas.  It was found that both models have
serious difficulties \citep{Jaaniste:1975, Komberg:1975}.  {
  Neutrinos were also considered, but rejected since they can form
  only supercluster-scale halos, about 1000 times more massive than
  halos around galaxies are.}
\begin{figure}[!]
\centering
\resizebox{0.45\textwidth}{!}
{\includegraphics{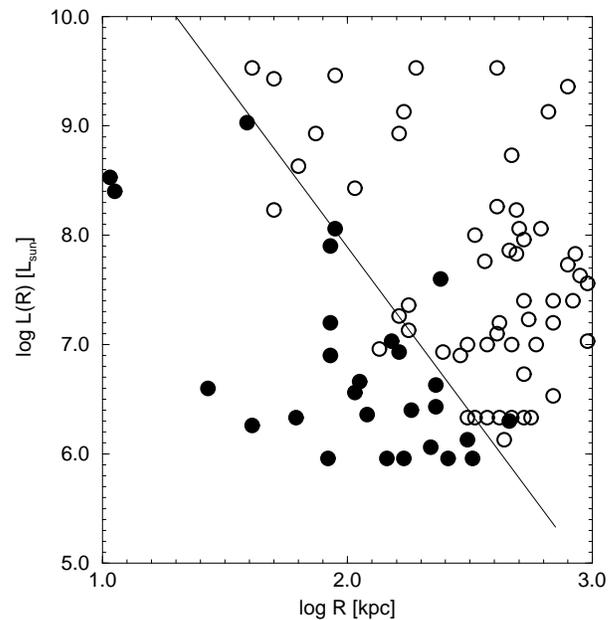}}
\caption{The distribution of luminosities of companion galaxies $L(R)$
  at various distances $R$ from the primary galaxy. Filled circles are
  for elliptical companions, open circles for spiral and irregular
  galaxies \citep{Einasto:1974b}.  Note the clear segregation of
  elliptical and spiral/irregular galaxies.}
\label{fig:segr}
\end{figure}

In Tbilisi the Third European Astronomical Meeting took place.  Here
the principal discussion was between the supporters of the classical
paradigm with conventional mass estimates of galaxies, and of the new
one with dark matter.  The major arguments supporting the classical
paradigm were summarized by \citet{Materne:1976}. Their most serious
argument was: {\em Big Bang nucleosynthesis suggests a low-density
  Universe with the density parameter $\Omega \approx 0.05$; the
  smoothness of the Hubble flow also favors a low-density Universe.}

It was clear that by sole discussion the presence and nature of dark
matter cannot be solved, new data and more detailed studies were
needed. The first very strong confirmation of the dark matter
hypothesis came from new extended rotation curves of galaxies.

\begin{figure}[ht]
\centering
\resizebox{0.45\textwidth}{!}
{\includegraphics{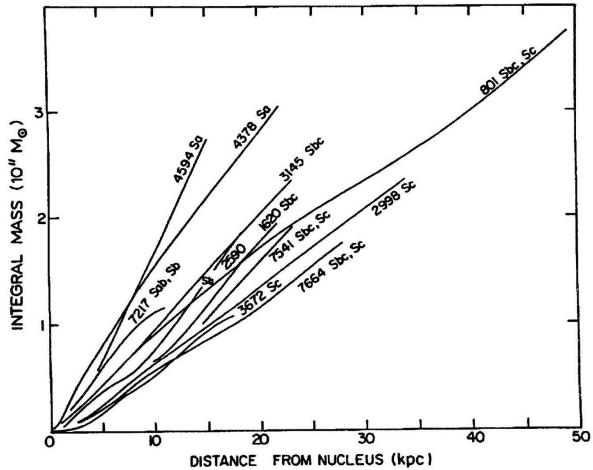}}
\caption{The integral masses as a function of the distance from the nucleus
  for spiral galaxies of various morphological type \citep{Rubin:1978}
  (reproduced by permission of the AAS and authors). }
\label{fig:Rubin}
\end{figure}

In early 1970s optical data on rotation of galaxies were available
only for inner bright regions of galaxies.  Radio observations of the
21-cm line reached much longer rotation curves well beyond the
Holmberg radius of galaxies.  All available rotation data were
summarized by \citet{Roberts:1975} in the IAU Symposium on Dynamics of
Stellar Systems held in Besancon (France) in September 1974.  Extended
rotation curves were available for 14 galaxies; for some galaxies data
were available until the galactocentric distance $\sim 40~h^{-1}$ kpc
(we use in this paper the Hubble constant in the units of $H_0 =
100~h$ km~s$^{-1}$~Mpc$^{-1}$), see Fig.~\ref{fig:M31} for M31.  About
half of galaxies had flat rotation curves, the rest had rotation
velocities that decreased slightly with distance.  In all galaxies the
local mass-to-light ratio in the periphery reached values over 100 in
Solar units.  To explain such high $M/L$ values Roberts assumed that
late-type dwarf stars dominate the peripheral regions.

\begin{figure*}[ht]
\centering
\resizebox{0.6\textwidth}{!}
{\includegraphics{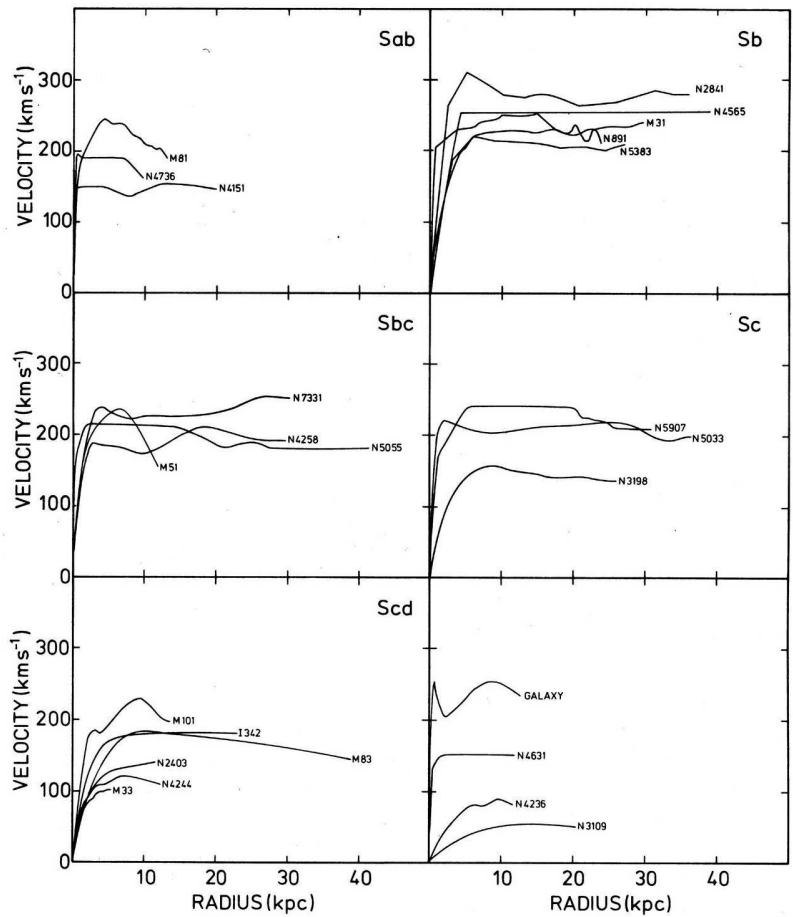}}
\caption{The rotation curves of spiral galaxies of various morphological type
  according to Westerbork radio observations \citep{Bosma:1978} (reproduced by
  permission of the author).  }
\label{fig:Bosma}
\end{figure*}

In mid-1970s Vera Rubin and her collaborators developed new sensitive
detectors to measure optically the rotation curves of galaxies at very
large galactocentric distances.  Their results suggested that
practically all spiral galaxies have extended flat rotation curves
\citep{Rubin:1978, Rubin:1980}.  The internal mass of galaxies rises
with distance almost linearly, up to the last measured point, see
Fig.~\ref{fig:Rubin}.

At the same time measurements of a number of spiral galaxies with the
Westerbork Synthesis Radio Telescope were completed, and mass
distribution models were built, all-together for 25 spiral galaxies
\citep{Bosma:1978}, see Fig.~\ref{fig:Bosma}.  Observations confirmed
the general trend that the mean rotation curves remain flat over the
whole observed range of distances from the center, up to $\sim 40$ kpc
for several galaxies. The internal mass within the radius $R$
increases over the whole distance interval.

These observational results confirmed the concept of the presence of dark
halos of galaxies with a high confidence.

Another very important measurement was made by Sandra Faber and
collaborators \citep{Faber:1976, Faber:1977, Faber:1979}.  They
measured the central velocity dispersions for 25 elliptical galaxies
and the rotation velocity of the Sombrero galaxy, a S0 galaxy with a
massive bulge and a very weak population of young stars and gas clouds
just outside the main body of the bulge.  Their data yielded for the
bulge of the Sombrero galaxy a mass-to-light ratio $M/L=3$, and for
the mean mass-to-light ratios for elliptical galaxies about 7, close
to the ratio for early type spiral galaxies. These observational data
confirmed estimates based on the calculations of physical evolution of
galaxies, made under the assumption that the lower mass limit of the
initial mass function (IMF) is for all galactic populations of the
order of 0.1~$M_\odot$.  These results showed that the mass-to-light
ratios of stellar populations in spiral and elliptical galaxies are
similar for a given color, and the ratios are much lower than those
accepted in earlier studies based on the dynamics of groups and
clusters.  In other words, high mass-to-light ratios of groups and
clusters of galaxies cannot be explained by visible galactic
populations.

Earlier suggestions on the presence of mass discrepancy in galaxies
and galaxy systems had been ignored by the astronomical community.
This time new results were taken seriously.  As noted by Kuhn, a
scientific revolution begins when leading scientists in the field
start to discuss the problem and arguments in favor of the new over
the old paradigm.

More data are slowly accumulating \citep{Sofue:2001}.  New HI
measurements from Westerbork extend the rotation curves up to 80 kpc
(galaxy UGC 2487) or even 100 kpc (UGC 9133 and UGC 11852) showing
flat rotation curves \citep{Noordermeer:2005}.  The HI distribution in
the Milky Way has been recently studied up to distances of 40 kpc by
\citet{Kalberla:2003, Kalberla:2007}.  The Milky Way rotation curve
has been determined by \citet{Xue:2008} up to $\sim 60 $ kpc from the
study of $\sim 2500$ Blue Horizontal Branch stars from SDSS survey,
and the rotation curves seems to be slightly falling from the 220 km
s$^{-1}$ value at the Sun location. Earlier determinations did not
extend so far and extrapolations were affected by the presence of the
ring-like structure in mass distribution at $\sim 14 $ kpc from the
center.  Implied values of the dark matter halo from different
measurements still differ between themselves by a factor 2 - 3, being
in the range from $10^{12} - 2.5 \times 10^{12} M_{\odot}$.  The
central density of dark matter halos of galaxies is surprisingly
constant, about $0.1~M_{\odot}$~pc$^{-3}$ \citep{Einasto:1974,
  Gilmore:2007}. Smallest dwarf galaxies have half-light radius about
120 pc, largest star clusters of similar absolute magnitude have
half-light radius up to 35 pc; this gap separates systems with and
without dark halos \citep{Gilmore:2008}.

\subsection{X-ray data on galaxies and clusters of galaxies}

Hot intra-cluster gas emitting X-rays was detected in almost all
nearby clusters and in many groups of galaxies by the Einstein X-ray
orbiting observatory.  Observations confirmed that the hot gas is in
hydrodynamical equilibrium, i.e. gas particles move in the general
gravitation field of the cluster with velocities which correspond to
the mass of the cluster \citep{Forman:1982, Sarazin:1988,
  Rosati:2002a}.

The distribution of the mass in clusters can be determined if the
density and the temperature of the intra-cluster gas are known.  This
method of determining the mass has a number of advantages over the use
of the virial theorem. First, the gas is a collisional fluid, and
particle velocities are isotropically distributed, which is not true
for galaxies as test particles of the cluster mass (uncertainties in
the velocity anisotropy of galaxies affect mass determinations).
Second, the hydrostatic method gives the mass as a function of radius,
rather than the total mass alone as given by the virial method.

Using Einstein X-ray satellite data the method was applied to
determine the mass of Coma, Perseus and Virgo clusters
\citep{Bahcall:1977, Mathews:1978a}.  The results were not very
accurate since the temperature profile was known only approximately.
The results confirmed previous estimates of masses made with the
virial method using galaxies as test particles.  The mass of the hot
gas itself is only about 0.1 of the total mass.  The luminous mass in
member galaxies is only a fraction of the X-ray emitting mass.

More recently clusters of galaxies have been observed in X-rays using
the ROSAT satellite (operated in 1990--1999), and the XMM-Newton and
Chandra observatories, launched both in 1999. The ROSAT satellite was
used to compile an all-sky catalog of X-ray clusters and galaxies.
More than 1000 clusters up to a redshift $\sim 0.5$ were
cataloged. Dark matter profiles have been determined in a number of
cases \citep{Humphrey:2006}.

The XMM and Chandra observatories allow us to get detailed images of
X-ray clusters, and to derive the density and temperature of the hot
gas \citep{Jordan:2004, Rasia:2006}.  Using the XMM observatory, a
survey of X-ray clusters was initiated to find a representative sample
of clusters at redshifts up to $z = 1$.  The comparison of cluster
properties at different redshifts allows the obtaining of more
accurate information on the evolution of clusters which depend
critically on the parameters of the cosmological model.

Chandra observations allow us to find the hot gas and total masses not only for
groups and clusters, but also for nearby galaxies (\citet{Humphrey:2006}, see
also \citet{Mathews:2006}, \citet{Lehmer:2008}).  For early-type (elliptical)
galaxies the virial masses found were 0.7--9$\times 10^{13}$~M$_\odot$. Local
mass-to-light ratio profiles are flat within an optical half-light radius
($R_{eff}$), rising more than an order of magnitude at $\sim 10 R_{eff}$,
which confirms the presence of dark matter.  The baryon fraction (most baryons
are in the hot X-ray emitting gas) in these galaxies is $f_b \sim 0.04 -
0.09$.  The gas mass profiles are similar to the profiles of dark matter
shifted to lower densities. The stellar mass-to-light ratios in these old
bulge dominated galaxies are $M_*/L_K \sim 0.5 - 1.9$ using the Salpeter IMF
(for the infrared K-band, the ratios for the B-band are approximately 4 times
higher). Interesting information of the chemical composition of the hot plasma in the halo of
the Milky Way were obtained in 2008 from the comparative study of the tiny
absorption lines in a few Galactic and extragalactic X-ray sources, giving the
total column density of O VII less than $5 \times 10^{15}$ cm$^{-2}$. Assuming
that the gaseous baryonic corona has the mass of order of $\sim 6 \times
10^{10} M_{\odot}$, as expected from the theory of galaxy formation, this
measurement implies a very low metallicity of the corona plasma, below
3.7 percent of the solar value.

\subsection{Galactic and extragalactic gravitational lensing}

Clusters, galaxies and even stars are so massive that their gravity bends and
focuses the light from distant galaxies, quasars and stars that lie far
behind.  There are three classes of gravitational lensing:

\begin{itemize}

\item{}  Strong lensing, where there are easily visible distortions such as the
formation of Einstein rings, arcs, and multiple images, see
Fig.~\ref{fig:a2218}. 

\item{} Weak lensing, where the distortions of background objects are much
  smaller and can only be detected by analyzing the shape distortions of a large
  number of objects.

\item{} Microlensing, where no shape distortion can be seen, but the amount
  of light received from a background object changes in time. The background
  source and the lens may be stars in the Milky Way or in nearby galaxies
  (M31, Magellanic Clouds). 

\end{itemize}

The strong lensing effect is observed in rich clusters, and allows us to
determine the distribution of the gravitating mass in clusters. Massive
galaxies can distort images of distant single objects, such as quasars: as a
result we observe multiple images of the same quasar. The masses of clusters
of galaxies determined using this method, confirm the results obtained by the
virial theorem and the X-ray data.

Weak lensing allows us to determine the distribution of dark matter in
clusters as well as in superclusters.  For the most luminous X-ray
cluster known, RXJ 1347.5-1145 at the redshift $z = 0.45$, the lensing
mass estimate is almost twice as high as that determined from the
X-ray data. The mass-to-light ratio is $M/L_B = 200 \pm 50$ in Solar
units \citep{Fischer:1997, Fischer:1997a}.  For other recent work on
weak lensing and X-ray clusters see
\citet{Bradac:2005,Dietrich:2005,Clowe:2006a, Massey:2007b}.

\begin{figure}[ht]
\centering
\resizebox{0.45\textwidth}{!}
{\includegraphics{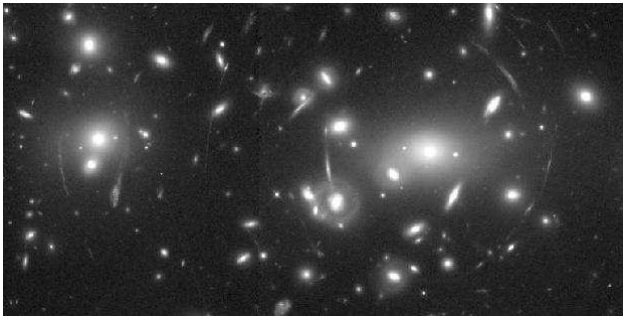}}
\caption{The Hubble Space Telescope image of the cluster Abell 2218. This
  cluster is so massive that its gravity bends the light of more distant
  background galaxies.  Images of background galaxies are distorted into
  stretched arcs (Astr. Pict. of the Day Jan. 11, 1998, Credit: W. Couch,
  R. Ellis). }
\label{fig:a2218}
\end{figure}

A fraction of the invisible baryonic matter can lie in small compact
objects -- brown dwarf stars or Jupiter-like objects.  To find the fraction of
these objects in the cosmic balance of matter, special studies have been
initiated, based on the microlensing effect.

Microlensing effects were used to find Massive Compact Halo Objects (MACHOs).
MACHOs are small objects as planets, dead stars (white dwarfs) or
brown dwarfs, which emit so little radiation that they are invisible most of
the time. A MACHO may be detected when it passes in front of a star and the
MACHOs gravity bends the light, causing the star to appear brighter.  Several
groups have used this method to search for the baryonic dark matter.  Some
authors claimed that up to 20 \% of dark matter in our Galaxy can be in
low-mass stars (white or brown dwarfs).  However, observations using the Hubble
Space Telescope's (HST) NICMOS instrument show that only about 6\% of the
stellar mass is composed of brown dwarfs. This corresponds to a negligible
fraction of the total matter content of the Universe \citep{Graff:1996a, 
Najita:2000}.

\section{The nature of Dark Matter}

By the end of 1970s most objections against the dark matter hypothesis
were rejected. In particular, luminous populations of galaxies have
found to have  lower mass-to-light ratios than expected previously,
thus the need of extra dark matter both in galaxies and clusters
is even stronger.  However, there remained three problems: 

\begin{itemize}
  
\item{} It was not clear how to explain the Big Bang nucleosynthesis
  constraint on the low density of matter, and the smoothness of the Hubble
  flow.
  
\item{} If the massive halo (corona) is not stellar nor gaseous, of what stuff
  is it made of?
  
\item{} And a more general question: in Nature everything has its
  purpose. If 90~\% of matter is dark, then there must be a reason for
  its presence. What is the role of dark matter in the history of the
  Universe?

\end{itemize}

First we shall discuss baryons as dark matter candidates.

\begin{figure}[ht]
\centering
\resizebox{0.48\textwidth}{!}
{\includegraphics{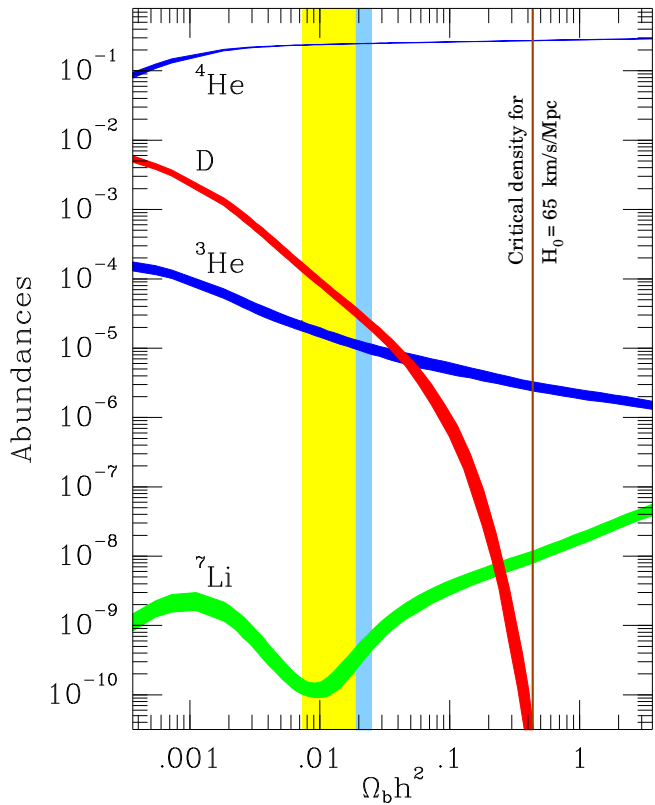}}
\caption{The big-bang production of the light elements.  The abundance of
  chemical elements is given as a function of the density of baryons,
  expressed in units of $\Omega_b h^2$ (horizontal axis).  Predicted abundances
  are in agreement with measured primeval abundances in a narrow range of
  baryon density \citep{Schramm:1998}.  }
\label{fig:Nsynt}
\end{figure}

\subsection{Nucleosynthesis constraints on the amount of baryonic matter}

According to the Big Bang model, the Universe began in an extremely hot and
dense state.  For the first second it was so hot that atomic nuclei could not
form -- space was filled with a hot soup of protons, neutrons, electrons,
photons and other short-lived particles.  Occasionally a proton and a neutron
collided and stuck together to form a nucleus of deuterium (a heavy isotope
of hydrogen), but at such high temperatures they were broken immediately by
high-energy photons \citep{Schramm:1998}.

When the Universe cooled off, these high-energy photons became rare enough that
it became possible for deuterium to survive.  These deuterium nuclei could keep
sticking to more protons and neutrons, forming nuclei of helium-3, helium-4,
lithium, and beryllium.  This process of element-formation is called
``nucleosynthesis''.  The denser proton and neutron ``gas'' is at this time, the
more of the total amount of light elements will be formed.  As the Universe expands,
 the density of protons and neutrons decreases and the process slows
down.  Neutrons are unstable (with a lifetime of about 15 minutes) unless they
are bound up inside a nucleus.  After a few minutes the free
neutrons will be gone and nucleosynthesis will stop.  There is only a small
window of time in which nucleosynthesis can take place, and the relationship
between the expansion rate of the Universe (related to the total matter/radiation 
density) and the density of protons and neutrons (the baryonic matter density)
determines how much of each of these light elements are formed in the early
Universe.

According to nucleosynthesis data baryonic matter makes up 0.04 of the
critical cosmological density, assuming $h \sim 0.7$ (Fig.~\ref{fig:Nsynt}).  Only a small
fraction, less than 10\%, of the baryonic matter is condensed to
visible stars, planets and other compact objects.  Most of the
baryonic matter is in the intergalactic matter, it is concentrated
also in hot X-ray coronas of galaxies and clusters.

\begin{figure}[ht]
\centering
\resizebox{0.4\textwidth}{!}
{\includegraphics{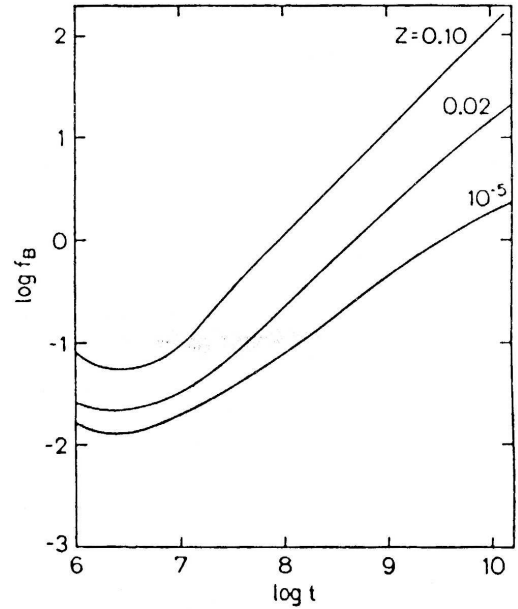}}
\caption{The evolution of mass-to-light ratios of galactic populations
  of different metal abundance $Z$ \citep{Einasto:1972}. }
\label{fig:MLevol}
\end{figure}

\subsection{Baryonic Dark Matter}

Models of the galaxy evolution are based on stellar evolution tracks, star
formation rates (as a function of time), and the initial mass function (IMF).
For IMF the \citet{Salpeter:1955} law is usually used:
\begin{equation}
F(m) = a m^{-n},
\label{salpeter}
\end{equation}
where $m$ is the mass of the forming star, and $a$ and $n$ are parameters.
This law cannot be used for stars of arbitrary mass, because in this case the
total mass of forming stars may be infinite.  Thus we assume that this law is
valid in the mass interval $m_0$ to $m_u$ (the lower and upper limit of the
forming stars, respectively).

{ Early models of physical evolution of galaxies were constructed
  by \citet{Tinsley:1968} and \citet{Einasto:1972}.  These models show
  that the mass-to-light ratio $M_i/L_i$ of the population $i$ depends
  critically on the lower limit of the IMF, $m_0$.  Calculations by
  \citet{Einasto:1972} show, that even for rather different physical
  conditions the value of $m_0$ changes only moderately
  (Fig.~\ref{fig:MLevol}).  An independent check of the correctness of
  the lower limit is provided by homogeneous stellar populations, such
  as star clusters.  Here we can assume that all stars were formed
  simultaneously, the age of the cluster can be estimated from the HR
  diagram, and the mass derived from the kinematics of stars in the
  cluster.  Such data are available for old metal-poor globular
  clusters, for relatively young medium-metal-rich open clusters, as
  well as for metal-rich cores of galaxies. } This check suggests that
in the first approximation for all populations similar lower mass
limits ($m_0 = 0.05 \dots 0.1~M_\odot$) can be used; in contracting
gas clouds above this limit the hydrogen starts burning, below not.
Using this mass lower limits we get for old metal-poor halo
populations $M_i/L_i \approx 1$, and for extremely metal-rich
populations in central regions of galaxies $M_i/L_i =10 \dots 100$, as
suggested by the central velocity dispersion in luminous elliptical
galaxies.  For intermediate populations (bulges and disks) one gets
$M_i/L_i = 3 \dots 10$, see Fig.~\ref{fig:MLevol}.  Modern data yield
for metal-rich populations lower values, due to more accurate
measurements of velocity dispersions in the central regions of
galaxies, as suggested in pioneering studies by \citet{Faber:1976,
  Faber:1977}, and more accurate input data for evolution models.

To get very high values of $M/L$, as suggested by the dynamics of
companion galaxies or rotation data in the periphery of galaxies, one
needs to use a very small value of the mass lower limit $m_0 \ll
10^{-3}~M_\odot$.  All known stellar populations have much lower
mass-to-light values, and form continuous sequences in color-$M/L$ and
velocity dispersion-$M/L$ diagrams.

For this reason it is very difficult to explain the physical and
kinematical properties of a stellar dark halo.  Dark halo stars form
an extended population around galaxies, and must have much higher
velocity dispersion than the stars belonging to the ordinary halo.  No
fast-moving stars as possible candidates for stellar dark halos were
found \citep{Jaaniste:1975}.  If the hypothetical population is of
stellar origin, it must be formed much earlier than all known
populations, because known stellar populations of different age and
metallicity form a continuous sequence of kinematical and physical
properties, and there is no place where to include this new population
into this sequence.  And, finally, it is known that star formation is
not an efficient process -- usually in a contracting gas cloud only
about 1~\%~ of the mass is converted to stars. Thus we have a problem
how to convert, in an early stage of the evolution of the Universe, a
large fraction of the primordial gas into this population of dark
stars. Numerical simulations suggest, that in the early universe only
a very small fraction of gas condenses to stars which ionize the
remaining gas and stop for a certain period further star formation
\citep{Cen:2003, Gao:2005}.

Stellar origin of dark matter in clusters was discussed by
\citet{Napier:1975}; they find that this is possible if the initial
mass function of stars is strongly biased toward very low-mass stars.
\citet{Thorstensen:1975} discussed the suggestion made by
\citet{Truran:1971} that there may have been a pre-galactic generation
of stars (population III), all of them more massive than the Sun,
which are now present as collapsed objects.  They conclude that the
total mass of this population is negligible, thus collapsed stars
cannot make up the dark matter.

{ Modern calculations suggests that metal-free population III stars
  are expected to be massive ($\sim 100$~M$_\odot$) due to large Jeans
  mass during the initial baryonic collapse (for a discussion see
  \citet{Reed:2005} and references therein). Thus population III stars
  are not suited to represent at the present epoch a high $M/L$ halo
  population.}

Recently weak stellar halos have been detected around several nearby
spiral galaxies at very large galactocentric distances. For instance,
a very weak stellar halo is found in M31 up to distance of 165 kpc
\citep{Gilbert:2006a,Kalirai:2006a}.  The stars of this halo have very
low metallicity, but have anomalously red color.  The total luminosity
and mass of these extended halos is, however, very small, thus these
halos cannot be identified with the dark halo.

Gaseous coronas of galaxies and clusters were discussed in 1970s by
\citet{Field:1972}, \citet{Silk:1974}, \citet{Tarter:1974},
\citet{Komberg:1975} and others.  The general conclusion from these
studies was that gaseous coronas of galaxies and clusters cannot
consist of neutral gas since the intergalactic hot gas would ionize
the coronal gas.  On the other hand, a corona consisting of hot
ionized gas would be observable.  Modern data show that a fraction of
the coronal matter around galaxies and in groups and clusters of
galaxies consists indeed of the X-ray emitting hot gas, but the amount
of this gas is not sufficient to explain the flat rotation curves of
galaxies \citep{Turner:2003}.

The results of  early discussions of the nature of dark halos were
inconclusive -- no appropriate candidate was found.  For many astronomers
this was an argument against the presence of dark halos.

\subsection{Non-baryonic Dark Matter and fluctuations of the CMB radiation}

Already in 1970s suggestions were made that some sort of non-baryonic
elementary particles, such as massive neutrinos, axions, photinos,
etc., may serve as candidates for dark matter particles.  There were
several reasons to search for non-baryonic particles as a dark matter
candidate.  First of all, no baryonic matter candidate did fit the
observational data.  Second, the total amount of dark matter is of the
order of 0.2--0.3 in units of the critical cosmological density, while
the nucleosynthesis constraints suggest that the amount of baryonic
matter cannot be higher than about 0.04 of the critical density.

A third very important observation was made which caused doubts to the
baryonic matter as the dark matter candidate. In 1964 Cosmic Microwave
Background (CMB) radiation was detected.  This discovery was a
powerful confirmation of the Big Bang theory.  Initially the Universe
was very hot and all density and temperature fluctuations of the
primordial soup were damped by very intense radiation; the gas was
ionized.  But as the Universe expanded, the gas cooled and at a
certain epoch called recombination the gas became neutral.  From this
time on, density fluctuations in the gas had a chance to grow by
gravitational instability.  Matter is attracted to the regions were
the density is higher and it flows away from low-density regions.  But
gravitational clustering is a very slow process.  Model calculations
show that in order to have time to build up all observed structures as
galaxies, clusters, and superclusters, the amplitude of initial
density fluctuations at the epoch of recombination must be of the
order of $10^{-3}$ of the density itself.  These calculations also
showed that density fluctuations are of the same order as temperature
fluctuations.  Thus astronomers started to search for temperature
fluctuations of the CMB radiation. None were found. As the accuracy of
measurement increased, lower and lower upper limits for the amplitude
of CMB fluctuations were obtained.  In late 1970s it was clear that
the upper limits are much lower than the theoretically predicted limit
$10^{-3}$ (see, for instance \citet{Parijskij:1978}).

\begin{figure*}[!]
\centering
\resizebox{0.48\textwidth}{!}
{\includegraphics{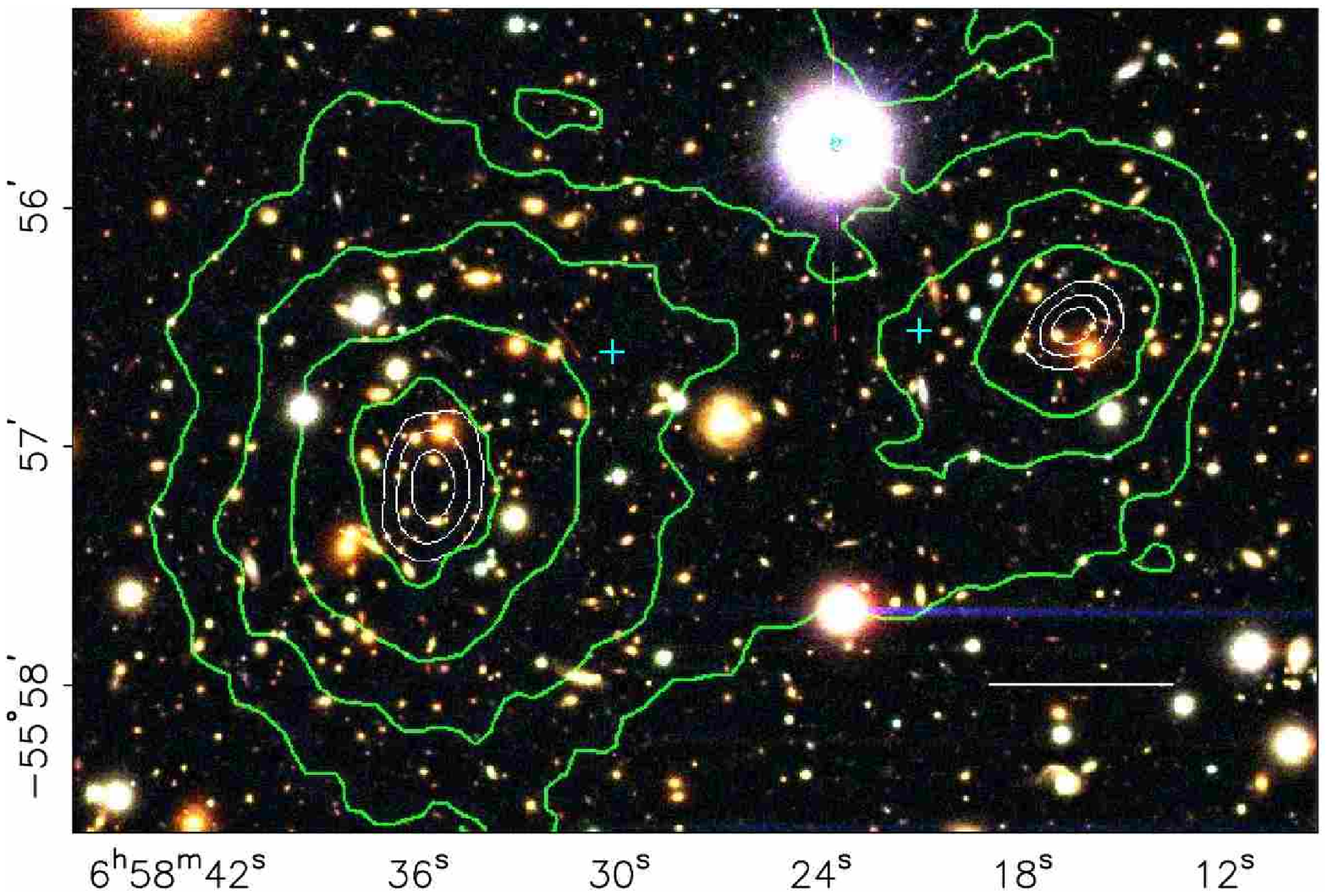}}
\resizebox{0.48\textwidth}{!}
{\includegraphics{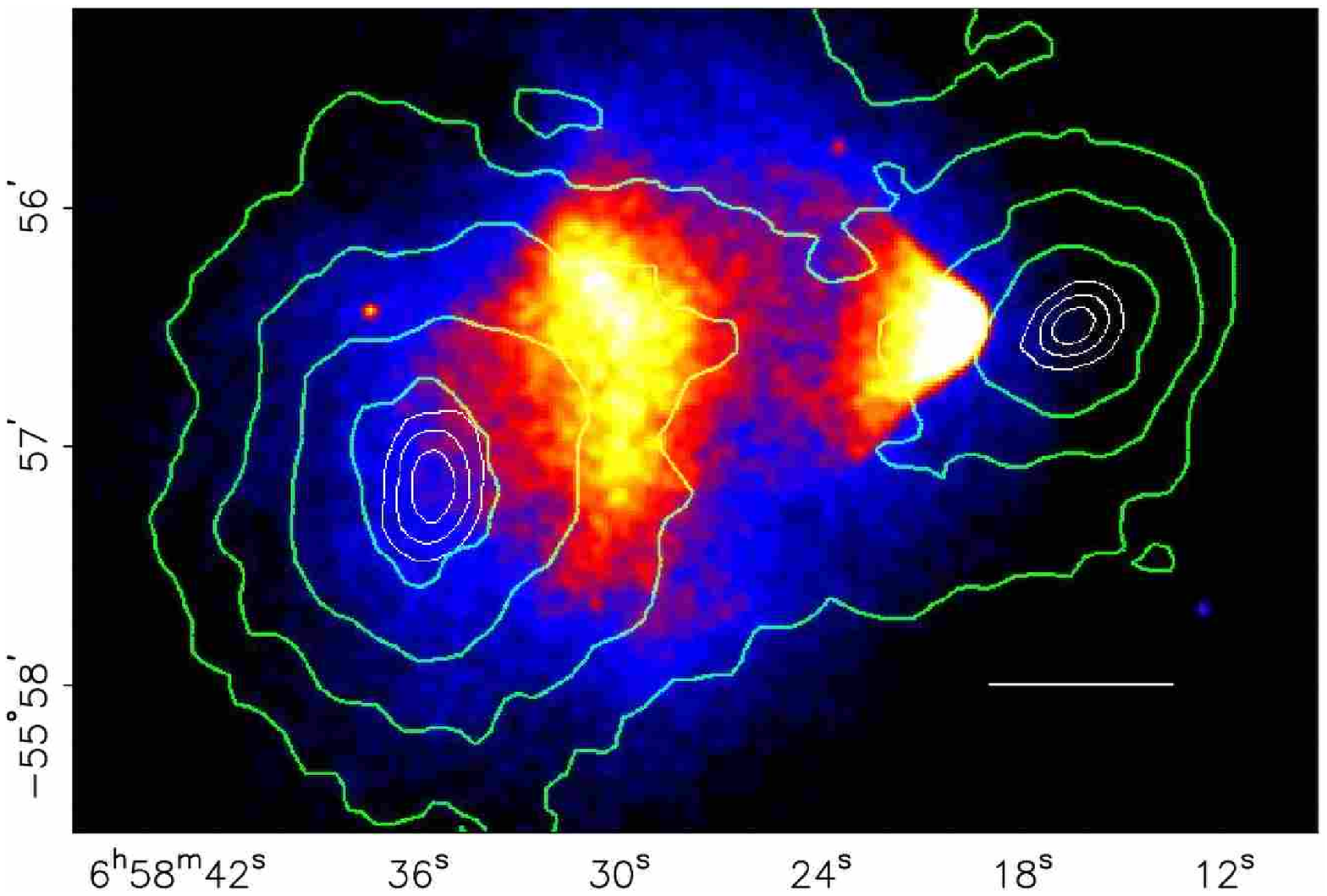}}
\caption{Images of the merging 'bullet' cluster 1E0657-558.  The left panel
  shows a direct image of the cluster obtained with the 6.5-m Magellan
  telescope in the Las Campanas Observatory, the right panel is a X-ray
  satellite Chandra image of the cluster. Shock waves of the gas are visible,
  the gas of the smaller 'bullet' cluster (right) lags behind the cluster
  galaxies. In both panels green contours are equidensity levels of the
  gravitational potential of the cluster, found using weak gravitational
  lensing of distant galaxies. The white bar has 200 kpc/h length at the
  distance of the cluster.  Note that contours of the gravitational potential
  coincide with the location of visible galaxies, but not with the location of
  the X-ray gas (the dominant baryonic component of clusters)
  \citep{Clowe:2006} (reproduced by permission of the AAS and authors).  }
\label{fig:bullet}
\end{figure*}

Then astronomers recalled the possible existence of non-baryonic
particles, such as heavy neutrinos. This suggestion was made
independently by several astronomers (\citet{Cowsik:1973, Szalay:1976,
  Tremaine:1979, Doroshkevich:1980b, Chernin:1981, Bond:1983b}) and
others.  They found that, if dark matter consists of heavy neutrinos,
then this helps to explain the paradox of small temperature
fluctuations of the cosmic microwave background radiation.  This
problem was discussed in a conference in Tallinn in April 1981.
Recent experiments by a Moscow physicist Lyubimov were announced,
which suggested that neutrinos have masses. If so, then the growth of
perturbations in a neutrino-dominated medium can start much earlier
than in a baryonic medium, and at the time of recombination
perturbations may have amplitudes large enough for structure
formation. The Lyubimov results were never confirmed, but it gave
cosmologists an impulse to take non-baryonic dark matter seriously.
In the conference banquet Zeldovich gave an enthusiastic speech: {\em
  ``Observers work hard in sleepless nights to collect data; theorists
  interpret observations, are often in error, correct their errors and
  try again; and there are only very rare moments of clarification.
  Today it is one of such rare moments when we have a holy feeling of
  understanding the secrets of Nature.''}  Non-baryonic dark matter is
needed to start structure formation early enough.  This example
illustrates well the attitude of theorists to new observational
discoveries -- the Eddington's test: {\em ``No experimental result
  should be believed until confirmed by theory''} (cited by
\citet{Turner:2000a}).  Dark matter condenses at early epoch and forms
potential wells, the baryonic matter flows into these wells and forms
galaxies \citep{White:1978}.

The search of dark matter can also be illustrated with the words of
Sherlock Holmes {\em ``When you have eliminated the impossible,
  whatever remains, however improbable, must be the truth''} (cited by
\citet{Binney:1987}).  The non-baryonic nature of dark matter explains
the role of dark matter in the evolution of the Universe, and the
discrepancy between the total cosmological density of matter and the
density of baryonic matter, as found from the nucleosynthesis
constraint.  Later studies have demonstrated that neutrinos are not
the best candidates for the non-baryonic dark matter, see below.

\subsection{Alternatives to Dark Matter}

The presence of large amounts of matter of unknown origin has given
rise to speculations on the validity of the Newton's law of gravity at
large distances.  One of such attempts is the Modified Newtonian
Dynamics (MOND), suggested by \citet{Milgrom:1987}, for a discussion
see \citet{Sanders:1990}.  Another attempt is the Modified Gravity
Theory (MOG), proposed by \citet{Moffat:2007}.  Indeed, MOND and MOG
are able to represent a number of observational data without assuming
the presence of some hidden matter.  However, there exist several
arguments which make these models unrealistic.

There exist numerous direct observations of the distribution of mass,
visible galaxies and the hot X-ray gas, which cannot be explained in
the MOND framework.  One of such examples is the ``bullet'' cluster 1E
0657-558 \citep{Clowe:2004, Markevitch:2004, Clowe:2006}, shown in
Fig.~\ref{fig:bullet}.  This is a pair of galaxy clusters, where the
smaller cluster (bullet) has passed the primary cluster almost
tangentially to the line of sight.  The hot X-ray gas has been
separated by ram pressure-stripping during the passage.  Weak
gravitational lensing yields the distribution of mass in the cluster
pair.  Lensing observations show that the distribution of matter is
identical with the distribution of galaxies.  The dominant population
of the baryonic mass is in X-ray gas which is well separated from the
distribution of mass.  This separation is only possible if the mass is
in the collisionless component, i.e. in the non-baryonic dark matter
halo, not in the baryonic X-ray gas.

\begin{figure*}[ht]
\centering
\resizebox{0.8\textwidth}{!}
{\includegraphics{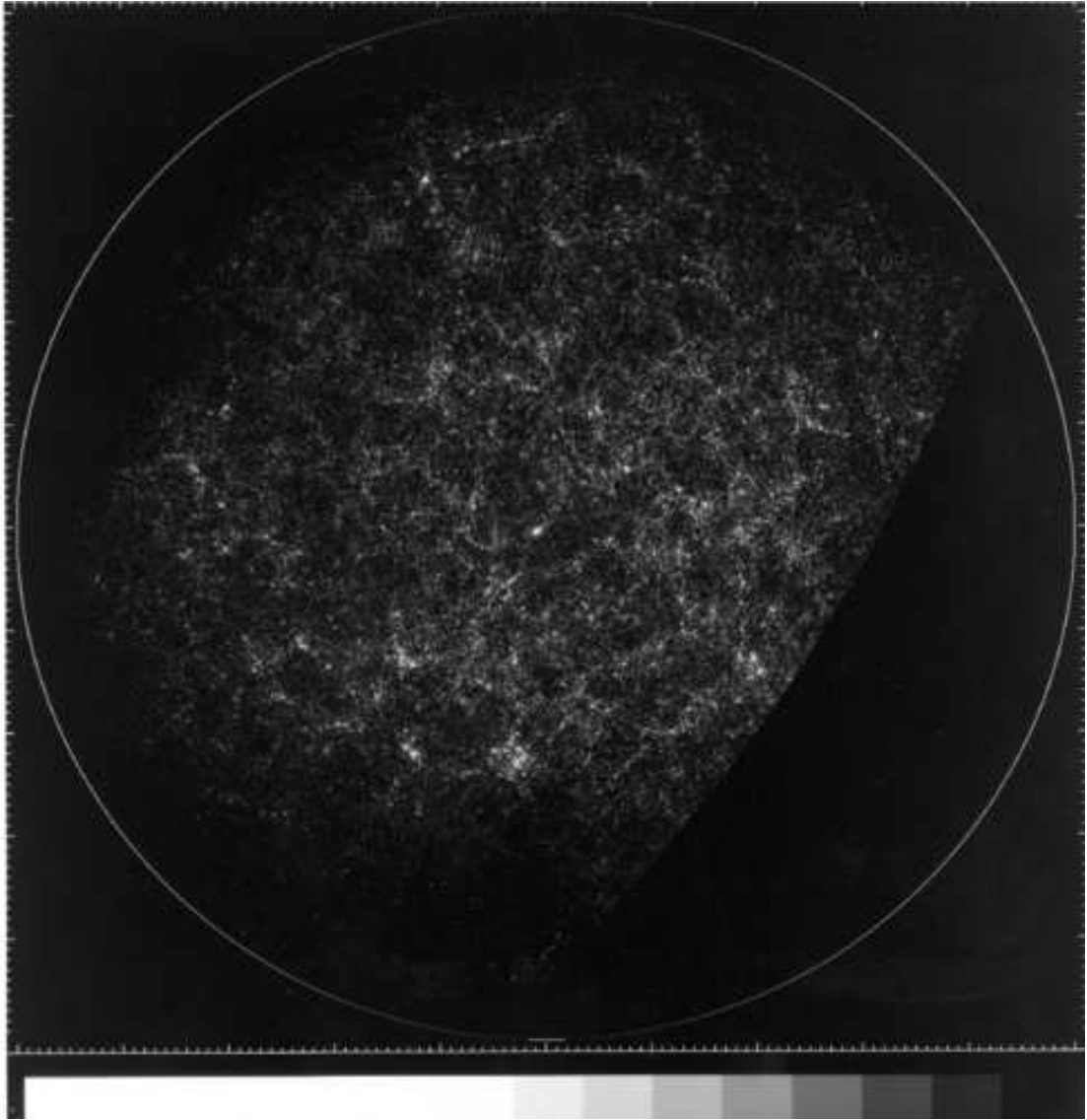}}
\caption{The two-dimensional distribution of galaxies according to the Lick
  counts \citep{Seldner:1977}. The north galactic pole is at the center, the
  galactic equator is at the edge. Superclusters are well seen, the Coma
  cluster is located near the center (reproduced by permission of the AAS and
  authors).  }
\label{fig:2dim}
\end{figure*}

{ Another example of a merging cluster is the rich cluster Cl
  0024+17 at redshift $z \approx 0.4$. This cluster has been observed
  using the Hubble Space Telescope Advanced Camera for Surveys
  (HST/ACS) data by \citet{Jee:2007wd}. The distribution of the hot
  intracluster medium (ICM) has been found using Chandra data.  Jee
  {\em et al.}  calculated the mass distribution, unifying both
  strong- and weak-lensing constraints. The mass reconstruction
  reveals a ringlike dark matter substructure at $r \sim 75"$, which
  surrounds a dense core at $r < 50"$. The redshift histogram of the
  cluster is bimodal, which is an indication of a high-speed
  line-of-sight collision of two massive clusters.  \citet{Jee:2007wd}
  interpret the mass distribution sub-structure as the result of the
  collision $\sim 1 - 2$ Gyr ago.  The formation of a ringlike
  structure is analogous to that in ring galaxies
  \citep{Lynds:1976bh}. N-body simulation of a collision of two
  massive clusters confirmed the formation of temporary ringlike
  structures before the final merging of clusters. Hot ICM forms by
  collision of two separate approximately isothermal clouds. The
  distribution of galaxies in both colliding clusters also remains
  almost unchanged during the passage. Further analysis of the
  distribution of ICM using data obtained with the HST ACS and the
  Subaru telescope by \citet{Jee:2010nx} has confirmed the peculiar
  dark matter structure in this cluster.  \citet{Jee:2007wd} conclude:
  {\em The ringlike mass structure surrounding the dense core not
    traced by the cluster ICM nor by the cluster galaxies serves as
    the most definitive evidence from gravitational lensing to date
    for the existence of dark matter. If there is no dark matter and
    the cluster ICM is the dominant source of gravity, the MONDian
    gravitational lensing mass should follow the ICM.}

A more general argument in favour of the presence of non-baryonic dark
matter comes from the CMB data. In the absence of large amounts of
non-baryonic matter during the radiation domination era of the
evolution of the Universe it would be impossible to get for the
relative amplitude of density fluctuations a value of the order of
$10^{-3}$, needed to form all observed structures.

It is fair to say that in comparison to the DM paradigm the
consequences of the various modifications of the Newtonian gravity,
that were mentioned above, have not been worked out in such a
detail. Thus it still needs to be seen if any of those modified
pictures could provide a viable alternative to the DM. However, one
has to keep in mind that despite us having a good idea of what might
make up a DM, the DM paradigm is remarkably simple: one just needs an
additional cold collisionless component that interacts only through
gravity. Once this component is accepted, a host of apparent problems,
starting from galaxy and galaxy cluster scales and extending to the
largest scales as probed by the large scale structure and CMB, get
miraculously solved. So in that respect one might say that there is
certainly some degree of elegance in the DM picture. On the other
hand, taking into account the simplicity of the DM paradigm, it is
quite hard to believe that any alternatives described above could
achieve a similar level of agreement with observational data over such
a large range of spatial and temporal scales. Indeed, it seems that
for different scales one might need ``a different MOND''.}

\section{Dark Matter and structure formation}

It is clear that if dark matter dominates in the matter budget of the
Universe, then the properties of dark matter particles determine the formation
and evolution of the structure of the Universe.  In this way the dark matter
problem is related to the large-scale structure of the Universe.

\subsection{The distribution of galaxies and clusters}

Already in the New General Catalogue (NGC) of nebulae, composed from
observations by William and John Herschel, a rich collection of nearby
galaxies in the Virgo constellation was
known. \citet{de-Vaucouleurs:1953} called this system the Local
Super-galaxy, presently it is known as the Virgo or Local
Supercluster.  Detailed investigation of the distribution of galaxies
became possible when Harlow Shapley started in the Harvard Observatory
a systematic photographic survey of galaxies in selected areas, up to
18th magnitude \citep{Shapley:1935, Shapley:1937, Shapley:1940}.
Shapley discovered several other rich superclusters, one of them is
presently named the Shapley Supercluster.  These studies showed also
that the {\em mean} spatial density of galaxies is approximately independent
of the distance and of the direction in the sky. In other words, the
Harvard survey indicated that galaxies are distributed in space
more-or-less homogeneously, as expected from the general cosmological
principle.

A complete photographic survey of galaxies was made in the Lick
Observatory with the 20-inch Carnegie astrograph by
\citet{Shane:1967}. Galaxy counts were made in cells of size $10'
\times 10'$, and the distribution of the number density of galaxies
was studied.  The general conclusion from this study was that galaxies
are mostly located in clusters, the number of galaxies per cluster
varying widely from pairs to very rich clusters of the Coma cluster
type.  The Lick counts were reduced by Jim Peebles and collaborators
to exclude count limit irregularities; the resulting distribution of
galaxies in the sky is shown in Fig.~\ref{fig:2dim}.

A much deeper photographic survey was made using the 48-inch Palomar
Schmidt telescope.  Fritz Zwicky used this survey to compile for the
Northern hemisphere a catalogue of galaxies and clusters of galaxies
\citep{Zwicky:1968}.  The galaxy catalogue is complete up to 15.5
photographic magnitude.  George Abell used the same survey to compile
a catalogue of rich clusters of galaxies for the Northern sky, later
the catalogue was continued to the Southern sky \citep{Abell:1958,
  Abell:1989}.  Using apparent magnitudes of galaxies approximate
distances (distance classes) were estimated for clusters in both
catalogues. Authors noticed that clusters of galaxies also show a
tendency of clustering, similar to galaxies which cluster to form
groups and clusters.  Abell called these second order clusters
superclusters, Zwicky -- clouds of galaxies.

The Lick counts as well as galaxy and cluster catalogues by Zwicky and
Abell were analyzed by Jim Peebles and collaborators
(\citet{Peebles:1973a}, \citet{Hauser:1973}, \citet{Peebles:1974a},
\citet{Peebles:1974b}).  To describe the distribution of galaxies
Peebles introduced the two-point correlation (or covariance) function
of galaxies \citep{Peebles:1975, Groth:1977, Fry:1978}.  This function
describes the probability to find a neighbor at a given angular
separation in the sky from a galaxy.  At small separations the spatial
galaxy correlation function can be approximated by a power law: $\xi =
(r/r_0)^{-\gamma}$, with the index $\gamma = 1.77 \pm 0.04$.  The
distance $r_0$, at which the correlation function equals unity, is
called the correlation length.  For galaxy samples its value is $r_0
\approx 5$~\Mpc, and for clusters of galaxies $r_0 \approx
30$~\Mpc. On scales $\geq 5$ times the correlation length the
correlation function is very close to zero, i.e. the distribution of
galaxies (clusters) is essentially random.

\begin{figure*}[ht]
\centering
\resizebox{0.45\textwidth}{!}
{\includegraphics{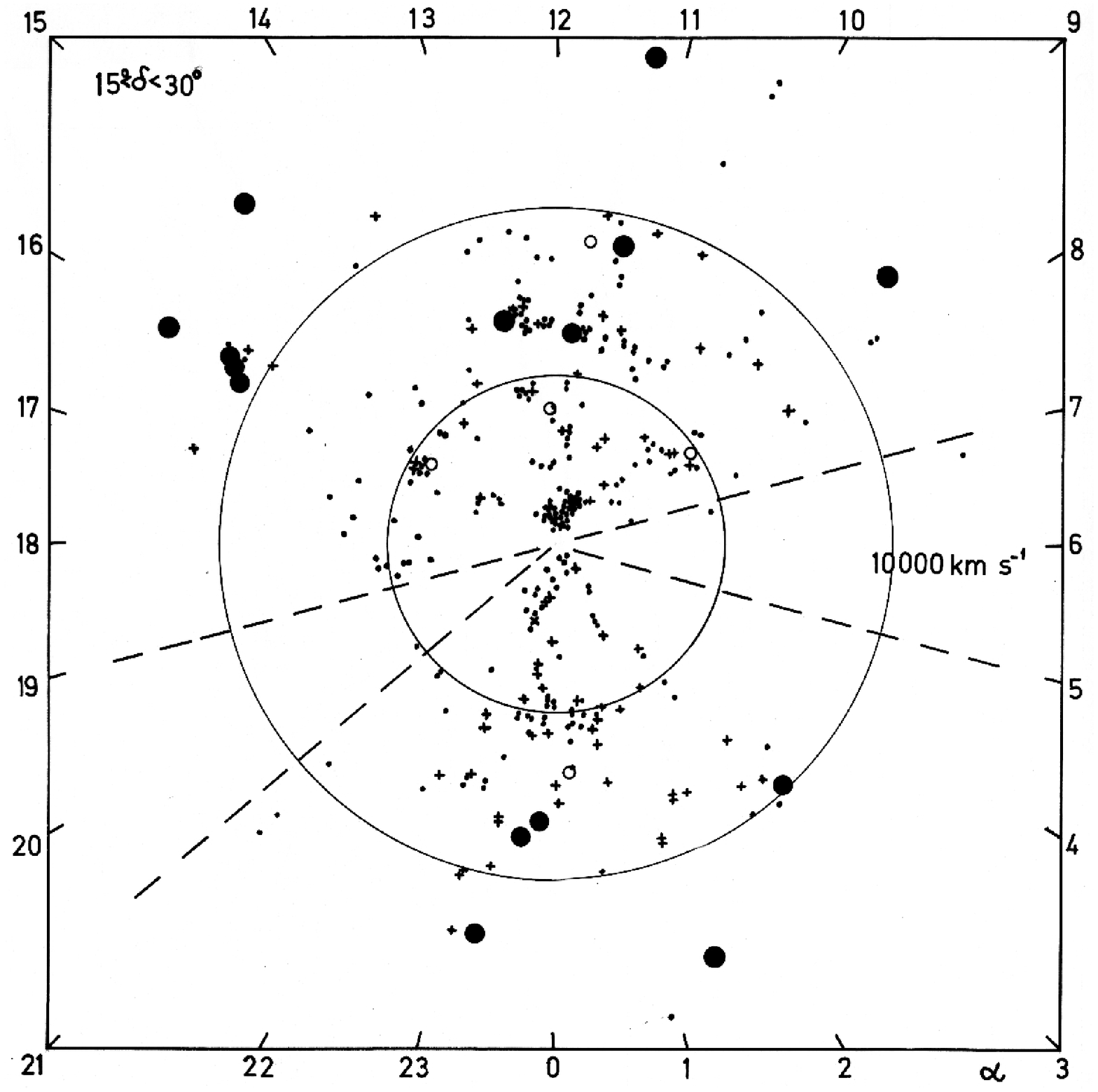}}
\resizebox{0.45\textwidth}{!}
{\includegraphics{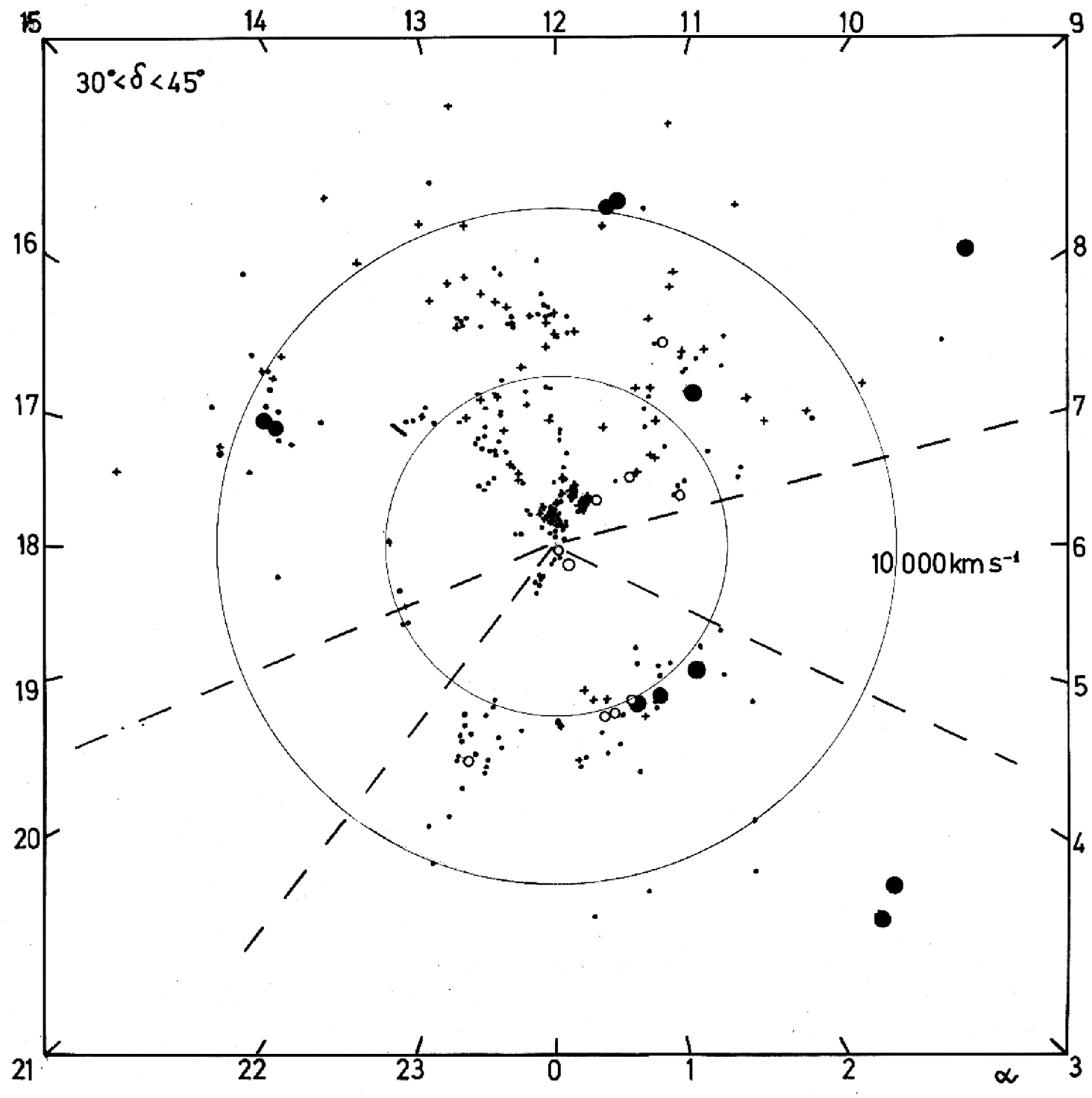}}
\caption{Wedge diagrams for two declination zones. Filled circles show rich
  clusters of galaxies, open circles -- groups, dots -- galaxies, crosses --
  Markarian galaxies. In the $15^\circ - 30^\circ$ zone two rich clusters at
  RA about 12 h are the main clusters of the Coma supercluster, in the
  $30^\circ - 45^\circ$ zone clusters at RA about 3 h belong to the main chain
  of clusters and galaxies of the Perseus-Pisces supercluster. Note the
  complete absence of galaxies in front of the Perseus-Pisces supercluster,
  and galaxy chains leading from the Local supercluster towards the
  Coma and Perseus-Oisces  superclusters \citep{Joeveer:1978a}.  }
\label{fig:wedges}
\end{figure*}

The conclusion from these studies, based on the apparent (2-dimensional)
distribution of galaxies and clusters in the sky confirmed the picture
suggested by \citet{Kiang:1967} and \citet{de-Vaucouleurs:1970}, among others,
that galaxies are hierarchically clustered.  However, this hierarchy does not
continue to very large scales as this contradicts observations, which show
that on very large scales the distribution is homogeneous.  A theoretical
explanation of this picture was given by Peebles in his hierarchical
clustering scenario of structure formation \citep{Peebles:1970, Peebles:1971}.

\subsection{Superclusters, filaments and voids}

In 1970s new sensitive detectors were developed which allowed the
measurement of redshifts of galaxies up to fainter magnitudes.  Taking
advance of this development several groups started to investigate the
environment of relatively rich clusters of galaxies, such as the Coma
cluster and clusters in the Hercules supercluster, with a limiting
magnitude about 15.5.  During this study Chincarini, Gregory, Rood,
Thompson and Tifft noticed that the main clusters of the Coma
supercluster, A1656 and A1367, are surrounded by numerous galaxies,
forming a cloud around clusters at the redshift $\sim$7000 km/s. The
Coma supercluster lies behind the Local supercluster, thus galaxies of
the Local supercluster also form a condensation in the same direction
at the redshift about 1000 km/s.  In between there is a group of
galaxies around NGC 4169 at the redshift $\sim$4000 km/s, and the
space between these systems is completely devoid of galaxies
\citep{Chincarini:1976, Gregory:1978}.  A similar picture was observed
in front of the Hercules and Perseus superclusters.

{ In 1970s there were two main rivaling theories of structure
  formation: the ``Moscow'' pancake theory by \citet{Zeldovich:1970},
  and the ``Princeton'' hierarchical clustering theory by
  \citet{Peebles:1971}.

  In developing the structure formation scenario Zeldovich used his
  previous experience in studying explosive phenomena -- he was a
  leading expert in the Soviet atomic bomb project.  He knew that in
  the early phase of the evolution of the Universe, when density
  fluctuations are very small, the {\em global} velocities, determined by
  the gravitational potential field, play the dominant role.  The
  development of the global velocity field leads to the formation of
  flat pancake-like systems.  In this scenario the structure forms
  top-down: first matter collects into pancakes and then fragments to
  form smaller units.

  The hierarchical clustering scenario is based on the Peebles
  experience of the study of galaxy clustering using Lick counts.  The
  clustering can be described by the correlation function, which
  describes the {\em local} clustering of galaxies.  According to this
  scenario the order of the formation of systems is the opposite:
  first small-scale systems (star-cluster sized objects) form, and by
  clustering systems of larger size (galaxies, clusters of galaxies)
  form; this is a bottom-up scenario.  }

{ In the Zeldovich team there were no experts on extragalactic
  astronomy, thus he asked Tartu astronomers for help in solving the question:
Can we find observational evidence which can be used to discriminate
between various theories of galaxy formation? } In solving the
Zeldovich question we started from the observational fact suggesting
that random velocities of galaxies are of the order of several hundred
km/s. Thus during the whole lifetime of the Universe galaxies have
moved from their place of origin only by about 1~\Mpc.  In other words
-- if there exist some regularities in the distribution of galaxies,
then these regularities must reflect the conditions in the Universe
during the formation of galaxies.

In mid-1970s first all-sky complete redshift surveys of galaxies were
just available: the \citet{de-Vaucouleurs:1976c} Second Revised
Catalogue of Galaxies, the Shapley-Adams revised catalogue by
\citet{Sandage:1981}, complete up to the magnitude 13.5 (new redshifts
were available earlier \citep{Sandage:1978}).  For nearby clusters of
galaxies and active (Markarian and radio) galaxies the redshift data
were also available. The common practice to visualize the
three-dimensional distribution of galaxies, groups and clusters of
galaxies is the use of wedge-diagrams. In these diagrams, where
galaxies as well as groups and clusters of galaxies were plotted, a
regularity was clearly seen: galaxies and clusters are concentrated to
identical essentially one-dimensional systems, and the space between
these systems is practically empty \citep{Joeveer:1978a}.  This
distribution was quite similar to the distribution of test particles
in a numerical simulation of the evolution of the structure of the
Universe prepared by the Zeldovich group (\citet{Doroshkevich:1980},
early results of simulation were available already in 1976).  In this
simulation a network of high- and low-density regions was seen:
high-density regions form cells which surround large under-dense
regions.  Thus the observed high-density regions could be identified
with Zeldovich pancakes { (for a detailed description of the search
  of regularities in galaxy distribution see \citet{Einasto:2001a})}.

The Large Scale Structure of the Universe was discussed at the IAU
symposium in Tallinn 1977, { following an initiative by Zeldovich}.
The amazing properties of the distribution of galaxies were reported
by four different groups: \citet{Tully:1978} for the Local
supercluster, \citet{Joeveer:1978a} for the Perseus supercluster,
\citet{Tarenghi:1978} for the Hercules supercluster, and
\citet{Tifft:1978} for the Perseus supercluster; see also
\citet{Gregory:1978} for the Coma supercluster and
\citet{Joeveer:1978} for the distribution of galaxies and clusters in
the Southern galactic hemisphere.  The presence of voids (holes) in
galaxy distribution was suggested in all four reports.  Tully \&
Fisher demonstrated a movie showing a filamentary distribution of
galaxies in the Local supercluster. 

J\~oeveer and Einasto emphasized the presence of fine structure:
groups and clusters of galaxies form chains in superclusters and
connect superclusters to a continuous network, as seen from wedge
diagrams in Fig.~\ref{fig:wedges}.  They demonstrated also
morphological properties of the structure of superclusters: clusters
and groups within the chain are elongated along the chain, and main
galaxies of clusters (supergiant galaxies of type cD) are also
elongated along the chain.  A long chain of clusters, groups and
galaxies of the Perseus-Pisces supercluster is located almost
perpendicular to the line of sight.  The scatters of positions of
clusters/groups along the chain in the radial (redshift) and
tangential directions are practically identical.  This demonstrates
that the chain is essentially an one-dimensional structure.

\begin{figure}[ht]
\centering
\resizebox{0.45\textwidth}{!}
{\includegraphics{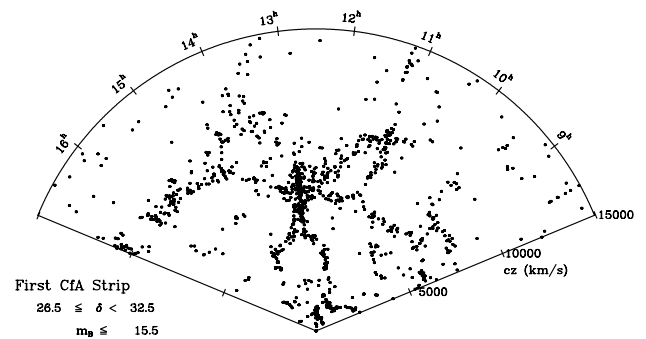}}
\caption{A slice of the Universe according to the CfA Second redshift
  survey \citep{de-Lapparent:1986}.  Galaxy chains
  connecting the Local and Coma superclusters are seen more clearly;
  the connection between the Hercules and Coma superclusters is also visible
  (reproduced by permission of the AAS and authors). 
}
\label{fig:cfa2}
\end{figure}

A direct consequence from this observation is that galaxies and
groups/clusters of the chain are already formed within the chain.  A
later inflow from random locations to the chain is excluded, since in
this case it would be impossible to stop galaxies and clusters in the
chain after the inflow.  The main results of the symposium were
summarized by Malcolm Longair as follows: {\em To me, some of the most
  exiting results presented at this symposium concerned the structure
  of the Universe on the largest scales.  Everyone seemed to agree
  about the existence of superclusters on scales $\sim 30 - 100$
  Mpc. But perhaps even more surprising are the great holes in the
  Universe. Peebles' picture, Einasto's analysis of the velocity
  distribution of galaxies which suggests a ``cell-structure'' and
  Tiffts's similar analysis argue that galaxies are found in
  interlocking chains over scales $\sim 50 - 100$ Mpc forming pattern
  similar to a lace-tablecloth.}

New data gave strong support to the pancake scenario by
\citet{Zeldovich:1978}.  However, some important differences between
the model and observations were evident.  First of all, numerical
simulations showed that there exists a rarefied population of test
particles in voids absent in real data. This was the first indication
for the presence of physical biasing in galaxy formation -- there is
primordial gas and dark matter in voids, but due to low density no
galaxy formation takes place here. Theoretical explanation of the
absence of galaxies in voids was given by Enn Saar
\citep{Einasto:1980}.  In over-dense regions the density increases
until the matter collapses to form compact objects (Zeldovich
pancakes).  In under-dense regions the density decreases
substantially, but never reaches a zero value -- gravity cannot
evacuate voids completely.

The second difference lies in the structure of galaxy systems in
high-density regions: in the original pancake model large-scale
structures (superclusters) have rather diffuse forms, real
superclusters consist of multiple intertwined filaments:
\citet{Joeveer:1978a}, \citet{Zeldovich:1982}, \citet{Oort:1983}.  In
the original pancake scenario small-scale perturbations were damped.
This scenario corresponds to the neutrino-dominated dark matter.
Neutrinos move with very high velocities which wash out small-scale
fluctuations.  Also, in the neutrino-dominated Universe superclusters
and galaxies within them form relatively late, but the age of old
stellar populations in galaxies suggests an early start of galaxy
formation, soon after the recombination epoch.  In other words, the
original pancake scenario was in trouble \citep{Bond:1982,
  Peebles:1982, Zeldovich:1982, Bond:1983, White:1983}.

The presence of voids in galaxy distribution was initially met with
skepticism, since 3-dimensional data were available only for bright
galaxies, and faint galaxies could fill voids.  However, independent
evidence was soon found.  A very large void was discovered in Bootes
by \citet{Kirshner:1981}. The filamentary nature of galaxy
distribution is very clearly seen in the 2nd Center for Astrophysics
(Harvard) Redshift Survey by Huchra, Geller and collaborators
\citep{de-Lapparent:1986, Huchra:1988}, complete up to 15.5 apparent
blue magnitude in the Northern Galactic hemisphere, see
Fig.~\ref{fig:cfa2}.

Huchra initiated a near-infrared survey of nearby galaxies, the Two
Micron All-Sky Survey (2MASS) \citep{Huchra:2000, Skrutskie:2006}.
Photometry in 3 near-infrared spectral bands is completed, it includes
about half a million galaxies up to the limiting K magnitude 13.5.
The redshifts are planned to be measured for all galaxies up to $K =
11.25$. The advantage of this survey is the coverage of low galactic
latitudes up to 5 degrees from the Galactic equator.  For the Southern
sky the redshift survey of 2MASS galaxies is almost completed using
the 6 degree Field Survey with the Australian large Schmidt telescope.
The filamentary character of the distribution of galaxies is very well
seen.

\begin{figure*}[ht]
\centering
\resizebox{0.8\textwidth}{!}
{\includegraphics{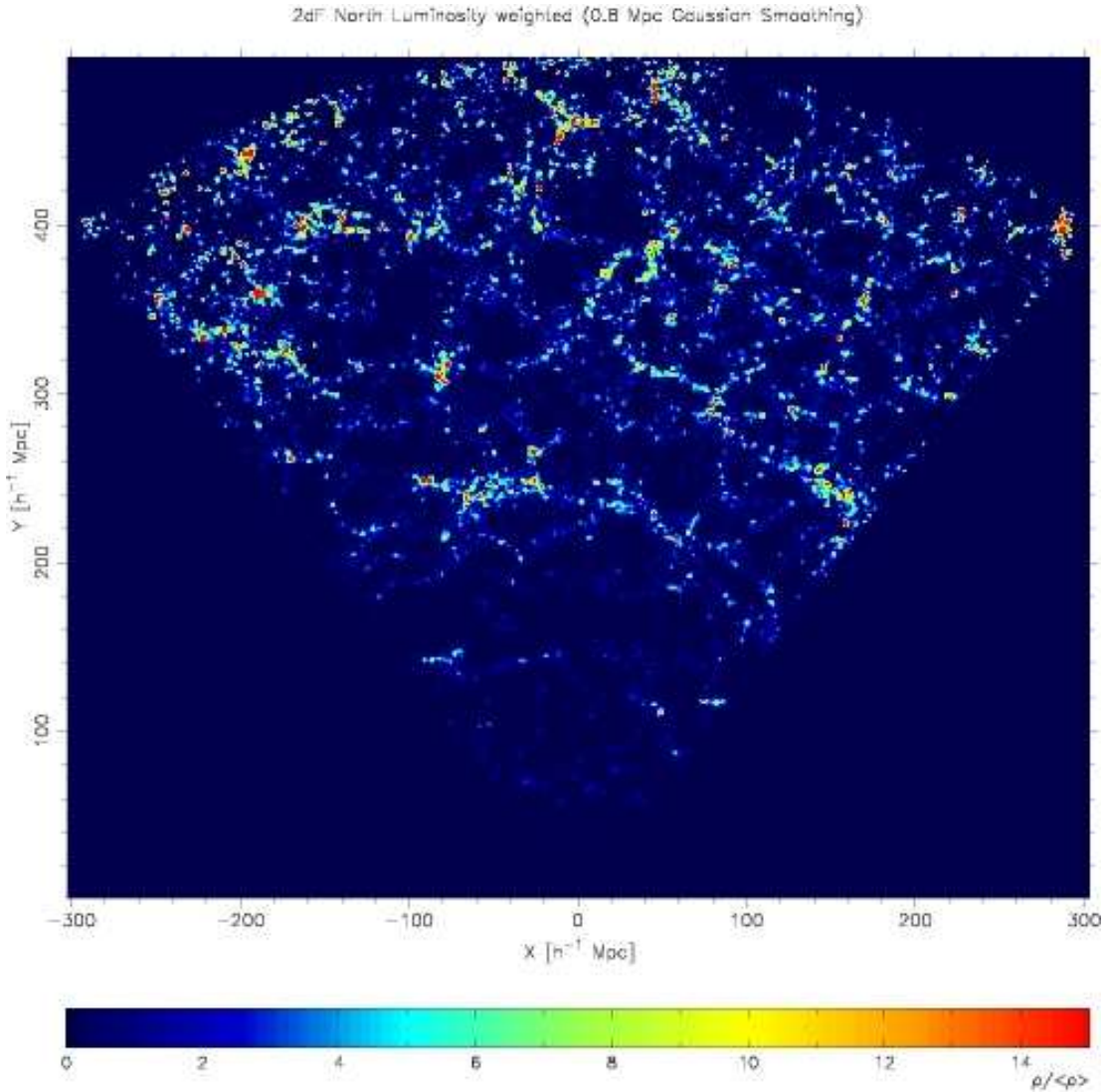}}
\caption{A wedge diagram of the luminosity density field of galaxies
  of the 2dFGRS Northern equatorial zone, $\pm 1.5$ degrees around the
  equator. The luminosity densities have been corrected for the
  incompleteness effect; the RA coordinate is shifted so that the plot
  is symmetrical around the vertical axis.  The rich supercluster at
  the distance $\sim 250$~\Mpc\ from the observer is SCL126, according
  to the catalogue by \citet{Einasto:2001}; it is the richest condensation in the   
  complex called Sloan Great Wall.  This figure
  illustrates the structure of the cosmic web (the supercluster-void
  network).  The density field shows that rich superclusters contain
  many rich clusters of galaxies, seen in picture as red dots.
  Filaments consisting of less luminous galaxies and located between
  superclusters and crossing large voids are also clearly seen \citep{Einasto:2007a}.  }
\label{fig:2dF}
\end{figure*}

A much deeper redshift survey up to the blue magnitude 19.4 was
recently completed using the Anglo-Australian 4-m telescope.  This Two
degree Field Galaxy Redshift Survey (2dFGRS) covers an equatorial
strip in the Northern Galactic hemisphere and a contiguous area in the
Southern hemisphere \citep{Colless:2001, Cross:2001}.  Over 250
thousand redshifts have been measured, which allows us to see and
measure the cosmic web (supercluster-void network) up to the redshift
0.2, corresponding to a co-moving distance about 575~\Mpc. The
luminosity density field calculated for the Northern equatorial slice
of the 2dFGRS is shown in Fig.~\ref{fig:2dF}.

Presently the largest project to map the Universe, the Sloan Digital
Sky Survey (SDSS) mentioned already before, has been initiated by a
number of American, Japanese, and European universities and
observatories \citep{York:2000, Stoughton:2002, Zehavi:2002,
  Abazajian:2009fj}.  The goal is to map a quarter of the entire sky:
to determine positions and photometric data in 5 spectral bands of
galaxies and quasars of about 100 million objects down to the red
magnitude {\tt r = 23}, and redshifts of all galaxies down to {\tt r =
  17.7} (about 1 million galaxies), as well as the redshifts of
Luminous Red Galaxies (LRG, mostly central galaxies of groups and
clusters) down to the absolute magnitude about $-20$.  All 7 data
releases have been made public. This has allowed the mapping of the
largest volume of the Universe so far. LRGs have a spatial density
about 10 times higher than rich Abell clusters of galaxies, which
allows us to sample the cosmic web with sufficient details up to a
redshift $\sim 0.5$.

\subsection{Structure formation in the Cold Dark Matter scenario}

A consistent picture of the structure formation in the Universe slowly
emerged from the advancement in the observational studies of the large
scale structure and the Cosmic Microwave Background, and in the
development of theory.  The most important steps along the way were
the following.

Motivated by the observational problems with neutrino dark matter, a
new dark matter scenario was suggested by \citet{Blumenthal:1982,
  Bond:1982, Pagels:1982, Peebles:1982, Bond:1983, Doroshkevich:1984}
with hypothetical particles as axions, gravitinos, photinos or
unstable neutrinos playing the role of dark matter. This model was
called the Cold Dark Matter (CDM) model, in contrast to the
neutrino-based Hot Dark Matter model. Newly suggested dark matter
particles move slowly, thus small-scale perturbations are not
suppressed, which allows an early start of the structure formation and
the formation of fine structure. Advantages of this model were
discussed by \citet{Blumenthal:1984}.

Next, the cosmological constant, $\Lambda$, was incorporated into the
scheme. Arguments favoring a model with the cosmological constant were
suggested already by \citet{Gunn:1975, Turner:1984, Kofman:1985}:
combined constraints on the density of the Universe, ages of galaxies,
and baryon nucleosynthesis.

Finally, there was a change in the understanding of the formation of
initial perturbations which later lead to the observed structure. To
explain the flatness of the Universe the inflation scenario was
suggested by \citet{Starobinsky:1980ys}, \citet{Guth:1981},
\citet{Starobinsky:1982ly}, \citet{Kofman:1985a} and others.
According to the inflation model in the first stage the expansion of
the Universe proceeded with accelerating rate (this early stage is
called the inflation epoch). Such an evolutionary scenario allows the
creation of the visible part of the Universe out of a small causally
connected region and explains why in the large scale the Universe
seems roughly uniform.  Perturbations of the field are generated by
small quantum fluctuations.  These perturbations form a Gaussian
random field, they are scale-invariant and have a purely adiabatic
primordial power spectrum \citep{Bardeen:1986fk}.
 
These ideas were progressively incorporated into the computer simulations of
increasing complexity.

Pioneering numerical simulations of the evolution of the structure of
the Universe were made in 1970s by \citet{Miller:1978},
\citet{Aarseth:1979} and the Zeldovich group
\citep{Doroshkevich:1980}, using direct numerical integration. In
early 1980s the Fourier transform was suggested to calculate the force
field which allowed the increase of the number of test particles.

Numerical simulations of structure evolution for the hot and cold dark
matter were compared by \citet{Melott:1983}, and by \citet{White:1983,
  White:1987} (standard CDM model with density parameter $\Omega_m =
1$).  In contrast to the HDM model, in the CDM scenario the structure
formation starts at an early epoch, and superclusters consist of a
network of small galaxy filaments, similar to the observed
distribution of galaxies.  Thus CDM simulations reproduce quite well
the observed structure with clusters, filaments and voids, including
quantitative characteristics (percolation or connectivity, the
multiplicity distribution of systems of galaxies,
\citet{Melott:1983}).
 
Models with the cosmological $\Lambda-$term were developed by
\citet{Gramann:1988}. Comparison of the SCDM and $\Lambda$CDM models
shows that the structure of the cosmic web is similar in both
models. However, in order to get the correct amplitude of density
fluctuations, the evolution of the SCDM model has to be stopped at an
earlier epoch.

The largest so far simulation of the evolution of the structure --
{\it the Millennium Simulation} -- was made in the Max-Planck
Institute for Astrophysics in Garching by Volker Springel and
collaborators \citep{Springel:2005, Gao:2005a, Springel:2006}.  The
simulation is assuming the $\Lambda$CDM initial power spectrum.  A
cube of the comoving size of 500~\Mpc\ was simulated using about 10
billion dark matter particles that allowed us to follow the evolution
of small-scale features in galaxies.  Using a semi-analytic model the
formation and evolution of galaxies was also simulated
\citep{Di-Matteo:2005, Gao:2005, Croton:2006}.  For simulated galaxies
photometric properties, masses, luminosities and sizes of principal
components (bulge, disk) were found.  The comparison of this simulated
galaxy catalogue with observations shows that the simulation was very
successful. The results of the Millennium Simulation are frequently
used as a starting point for further more detailed simulations of
evolution of single galaxies.
 
One difficulty of the original pancake scenario by Zeldovich is the
shape of objects formed during the collapse.  It was assumed that
forming systems are flat pancake-like objects, whereas dominant
features of the cosmic web are filaments.  This discrepancy has found
an explanation by \citet{Bond:1996}.  They showed that in most cases
just essentially one-dimensional structures, i.e. filaments form.

The $\Lambda$CDM model of structure formation and evolution combines
all essential aspects of the original structure formation models, the
pancake and the hierarchical clustering scenario.  First structures
form at very early epochs soon after the recombination in places where
the primordial matter has the highest density.  This occurs in the
central regions of future superclusters.  First objects to form are
small dwarf galaxies, which grow by infall of primordial matter and
other small galaxies.  Thus, soon after the formation of the central
galaxy other galaxies fall into the gravitational potential well of
the supercluster.  These clusters have had many merger events and have
``eaten'' all its nearby companions. During each merger event the
cluster suffers a slight shift of its position. As merger galaxies
come from all directions, the cluster sets more and more accurately to
the center of the gravitational well of the supercluster. This
explains the fact that very rich clusters have almost no residual
motion in respect to the smooth Hubble flow. Numerous examples of the
galaxy mergers are seen in the images of galaxies collected by the
Hubble Space Telescope, see Fig.~\ref{fig:bullet}.

\subsection{The density distribution of Dark Matter}

{ Flat rotation curves of galaxies suggest, that the radial density
  distribution in galaxies, including stellar populations,
  interstellar gas and dark matter, is approximately isothermal:
  $\rho(r) \sim r^{-2}$.  As the dark matter is the dominating
  population, its density profile should also be close to an
  isothermal sphere.  Thus, in the first approximation, one can use
  for the dark matter population a pseudo-isothermal profile
\begin{equation}
\rho(a) = {\rho(0)  \over  1+(a/a_0)^2},
\label{isotherm}
\end{equation}
where $a$ is the semi-major axis of the isodensity ellipsoid, and
$a_0$ is the effective radius of the population, called also the core
radius.  This law cannot be used at very large distances from the
center, $a$, since in this case the mass of the population would be
infinite.  This density law is an example of so-called cored profiles.

In the early 1990s, the results of high-resolution numerical N-body
simulations of dark matter halos based on the collisionless CDM model
became available. The simulations did not show the core-like behaviour
in the inner halos, but were better described by a power-law density
distribution, the so-called cusp.  \citet{Navarro:1997dq} investigated
systematically simulated DM halos for many different sets of
cosmological parameters. They found that the whole mass density
distribution could be well described by an ``universal density
profile''
\begin{equation}
\rho(a) = {\rho_i  \over  (a/a_s)(1+a/a_s)^2},
\label{NFW}
\end{equation}
where $\rho_i$ is related to the density of the universe at the time
of the halo collapse, and $a_s$ is the characteristic radius
(semi-major axis) of the halo. This profile, known as the ``NFW
profile'', cannot be applied to the very center of the halo, since in
this case the density would be infinite. Near the center of the halo
the density rises sharply, forming a ``cusp''.

The ``core-cusp problem'' has been a subject of many recent studies,
based both on observational data as well as on results of very
high-resolution numerical simulations. A review of these efforts is
given by \citet{de-Blok:2010ly}.  To find the DM-halo density profile
\citet{de-Blok:2010ly} used a collection of HI rotation curves of
dwarf galaxies, which are dominated by dark matter. To get a better
resolution near the center H$_\alpha$ long-slit rotation curves were
analysed. These rotation curves indicate the presence of
constant-density or mildly cuspy dark matter cores. Strongly cuspy
central profiles as the NFW one are definitely excluded.

\citet{Navarro:2010vn} performed a detailed numerical study of the
distribution of the mass and velocity dispersion of DM halos in the
framework of the {\em Aquarius Project}. The formation and evolution
of 6 different galaxy-sized halos were simulated several times at
varying numerical resolution, the highest resolution simulation had up
to 4.4 billion particles per halo. Authors find that the mass profiles
of halos are best represented by the Einasto profile
(\ref{explaw}). The radial dependence of the inner logarithmic slope,
$\gamma(r) = d \ln \rho(r)/d \ln r$ follows a power law, $\gamma(r)
\sim r^\alpha$.  The shape parameter $\alpha$ varies slightly from
halo to halo.  \citet{Navarro:2010vn} calculated also the
pseudo-phase-space density, $Q(r) = \rho(r)/\sigma^3(r)$, where
$\sigma^2(r)$ is the mean squared velocity dispersion, averaged in a
spherical shell of radius $r$. Authors find that pseudo-phase-density
profiles of all halos follow an identical power law, $Q(r) \sim
r^{-1.875}$. The origin of this behaviour is unclear, but its
similarity for all halos may reflect a fundamental structural property
of DM halos.

On cluster and supercluster scales the distribution of dark matter can
be most accurately found by gravitational lensing.
\citet{Gavazzi:2003ai} used strong lensing data obtained with the ESO
Very Large Telescope to analyse the radial mass profile of the galaxy
cluster MS 2137.3-2353 at redshift $z=1.6$. The mass density can be
represented both with the isothermal model as well as the NFW model.

\citet{Massey:2007b} used 575 pointings of the HST SCS Wide Field
Camera to cover a region of 1.637 square degrees, and measured shapes
of half a million distant galaxies to calculate the density
distribution around a rich cluster of galaxies at redshift $z=0.73$.
Additional information was obtained using the XMM-Newton X-ray
satellite observations, and distribution of galaxies in and around the
cluster. The most prominent peak in the distribution of matter using
all tracers is the cluster itself. Weak lensing shows additionally the
3-dimensional distribution of matter around the cluster. The
filamentary character of the mass distribution is clearly seen, the
cluster lies at the connection of several filaments.

\citet{Heymans:2008tg} applied weak lensing analysis of a HST STAGES
survey to reconstruct dark matter distribution in the $z=0.165$ Abell
901/902 supercluster. Authors detect the four main structures of the
supercluster. The distribution of dark matter is well traced by the
cluster galaxies. The high number density of HST data allows us to
produce a density map with sub-arcminute resolution. This allowed us
to resolve the morphology of dark matter structures. Profiles of DM
are far from the spherically symmetric NFW models. An extension of the
dark matter distribution is in the direction of an in-falling X-ray
group Abell 901$\alpha$, showing the filamentary character of the
distribution of dark matter.

\citet{Humphrey:2006,Humphrey:2010ve} used Chandra X-ray observatory
data to investigate the mass profiles of samples of 7 and 10 galaxies,
groups and clusters, respectively, spanning about 2 orders of
magnitude in virial mass. They find that the {\em total} as well as
{\em DM} mass density distributions can be well represented by a
NFW/Einasto profile. For the projected density of baryonic stellar
populations they use the \citet{Sersic:1968dz} profile. As shown by
\citet{Einasto:1965,Einasto:1974a}, the generalized exponential
expression (\ref{explaw}) can be applied as well for the spatial
density of galactic population. Thus the \citet{Humphrey:2010ve} study
suggests that an identical expression (\ref{explaw}) (Einasto profile)
can be applied for the total mass density distribution, the baryonic
and the DM mass density distribution.

The coincidence is remarkable, since the fraction of baryonic matter
in the total mass distribution in clusters varies with radius
considerably. This ``galaxy-halo conspiracy'' is similar to that which
establishes flat rotation curves in galaxies -- the ``bulge-halo
conspiracy''.  These coincidences suggest the presence of some sort of
interaction between the dominating stellar population (bulge) and the
dark matter halo, both on galactic and cluster scales.  We note that
an analogous relation exists between the mass of the central black
hole and the velocity dispersion of the bulge of elliptical galaxies
(see \citet{Gultekin:2009hc} for a review). Another interesting
phenomenon is the almost constant central density of DM halos, as
noted by \citet{Einasto:1974, Gilmore:2007} and \citet{Donato:2009gb}.

These studies demonstrate that the density enhancements of the dark
matter have a structure, which is very similar to the distribution of
galaxies: both forms of matter follow the same pattern of the cosmic
web, as expected \citep{Bond:1996}. }

\section{Matter-energy content of the Universe}

\subsection{Dark Matter and Dark Energy} 

In early papers on dark matter the total density due to visible and
dark matter was estimated to be about 0.2 of the critical cosmological
density.  These estimates were based on the dynamics of galaxies in
groups and clusters.  This density estimate can be interpreted in two
different ways: either we live in an open Universe where the total
density is less than the critical density, or there exists some
additional form of matter/energy which allows the Universe to be flat,
i.e. to have the critical total density.  The additional term was
identified with the Einstein $\Lambda$-term, so that the total
matter/energy density was taken to be equal to the critical
cosmological density (\citet{Gunn:1975, Turner:1984,
  Kofman:1985}). Initially there was no direct observational evidence
in favor of this solution and it was supported basically on general
theoretical grounds.  In its early evolution the size of the Universe
increases very rapidly and any deviation from the exact critical
density would lead to a rapid change of the relative density, either
to zero, if the initial density was a bit less than the critical one,
or to infinity, if it was greater than critical.  In other words, some
fine tuning is needed to keep the density at all times equal to the
critical one. The fine tuning problem can be eliminated if one assumes
an early accelerated epoch of expansion of the Universe (inflation).

\begin{figure}[ht]
\centering
\resizebox{0.49\textwidth}{!}
{\includegraphics{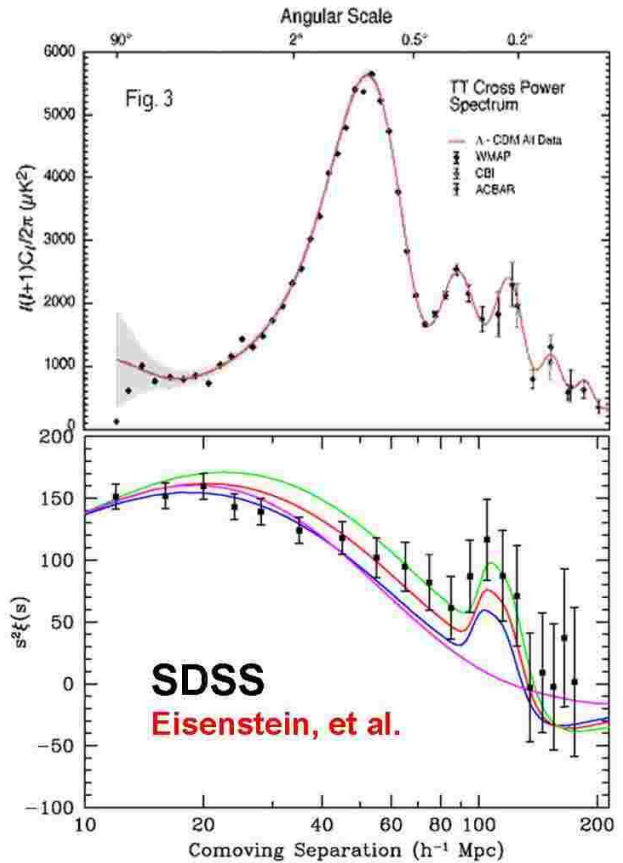}}
\caption{Upper panel shows the acoustic peaks in the angular power spectrum of
  the CMB radiation according to the WMAP and other recent data, compared with
  the $\Lambda$CDM model using all available data.  The lower panel shows the
  signature of baryonic acoustic oscillations in the matter two-point
  correlation function \citep{Eisenstein:2005, Kolb:2007}
 (reproduced by permission of the author)}.
\label{fig:WMAPspectr}
\end{figure}

In subsequent years several new independent methods were applied to estimate
the cosmological parameters.  Of these new methods two desire special
attention.  One of them is based on the measurements of small fluctuations of
the Cosmic Microwave Background (CMB) radiation, and the other on the
observation of distant supernovae.

According to the present cosmological paradigm the Universe was initially very
hot and ionized.  The photons provided high pressure and prevented baryons
to cluster.  Perturbations of baryons did not grow, but oscillated as sound
waves. The largest possible amplitude of these oscillations is at the wavelength  equal 
to the sound horizon size at the decoupling.  This wavelength is seen as the first
maximum in the angular power spectrum of the CMB radiation.  The following
maxima correspond to overtones of the first one.  The fluctuations of CMB
radiation were first detected by the COBE satellite.  The first CMB data were
not very accurate, since fluctuations are very small, of the order of
$10^{-5}$.  Subsequent experiments carried out using balloons, ground based
instruments, and more recently the Wilkinson Microwave Anisotropy Probe (WMAP)
satellite, allowed the measurement of the CMB radiation and its power spectrum with a
much higher precision \citep{Spergel:2003}.
 The position of the first maximum of the power
spectrum depends on the total matter/energy density. Observations confirm the
theoretically favored value 1 in units of the critical cosmological density,
see Fig.~\ref{fig:WMAPspectr}.

The small initial overdensities of the primordial cosmic medium launch
shock waves in the photon-baryon fluid.  After some time photons
completely decouple from baryons, and the baryons loose photon
pressure support. The shock stops after traveling a distance of about
150 Mpc (in comoving coordinates).  This leads to an overdensity of
the baryonic medium on a distance scale of 150 Mpc.  This overdensity
has been recently detected in the correlation function of Luminous Red
Galaxies of the SDSS survey \citep{Eisenstein:2005, Hutsi:2006a}, see
lower panel of Fig.~\ref{fig:WMAPspectr}.  Baryonic acoustic
oscillations depend on both the total matter/energy density and the
baryon density, thus allowing us to estimate these parameters.

Another independent source of information on cosmological parameters
comes from the distant supernova experiments.  Two teams, led by
\citet{Riess:1998, Riess:2007} (High-Z Supernova Search Team) and
\citet{Perlmutter:1999} (Supernova Cosmology Project), initiated
programs to detect distant type Ia supernovae in the early stage of
their evolution, and to investigate with large telescopes their
properties. These supernovae have an almost constant intrinsic
brightness (depending slightly on their evolution). By comparing the
luminosities and redshifts of nearby and distant supernovae it is
possible to calculate how fast the Universe was expanding at different
times. The supernova observations give strong support to the
cosmological model with the $\Lambda$ term, see Fig.~\ref{fig:DSP}.

\begin{figure}[ht]
\centering
\resizebox{0.45\textwidth}{!}
{\includegraphics{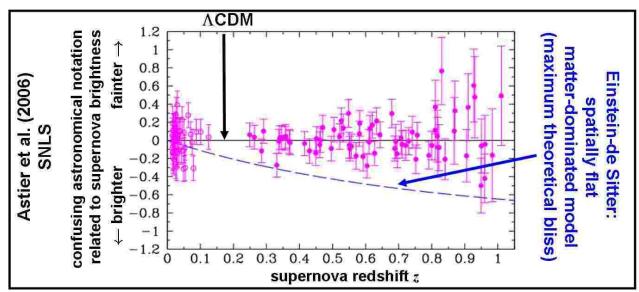}}
\caption{Results of the Supernova Legacy Survey: apparent magnitudes of
  supernovae are normalized to the standard $\Lambda$CDM model, shown as
  solid line. Dashed line shows the Einstein-de Sitter model with $\Omega_m =
  1$ \citep{Kolb:2007}  (reproduced by permission of the author). 
}
\label{fig:DSP}
\end{figure}

Different types of dark energy affect the rate at which the Universe
expands, depending on their effective equation of state. The
cosmological constant is equivalent to the vacuum.  The other possible
candidate of dark energy is quintessence (a scalar field) that has a
different and generally variable equation of state.  Each variant of
dark energy has its own equation of state that produces a signature in
the Hubble diagram of the type Ia supernovae \citep{Turner:2000a,
  Turner:2003}.

The combination of the CMB and supernova data allows us to estimate
independently the matter density and the density due to dark energy,
shown in Fig.~\ref{fig:DSP_CMB}. The results of this combined approach
imply that the Universe is expanding at an accelerating rate.  The
acceleration is due to the existence of some previously unknown dark
energy (or cosmological constant) which acts as a repulsive force (for
reviews see \citet{Bahcall:1999}, \citet{Frieman:2008}).

\begin{figure}[ht]
\centering
\resizebox{0.49\textwidth}{!}
{\includegraphics{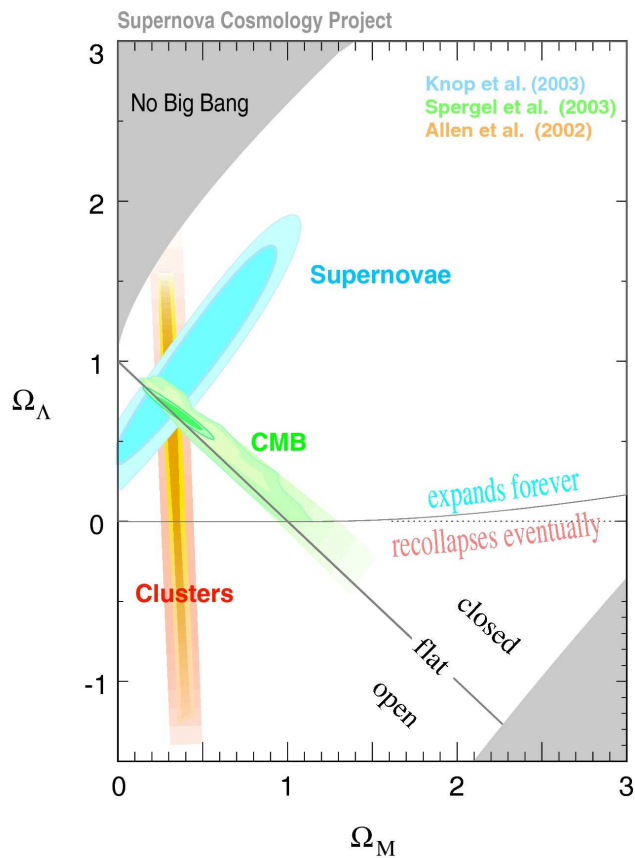}}
\caption{Combined constraints to cosmological densities
  $\Omega_\Lambda$ and $\Omega_M$, using supernovae, CMB and cluster
  abundance data.  The flat Universe with $\Omega_\Lambda + \Omega_M =
  1$ is shown with solid line \citep{Knop:2003}.  }
\label{fig:DSP_CMB}
\end{figure}

Independently, the matter density parameter has been determined from
clustering of galaxies in the 2-degree Field Redshift Survey and the
Sloan Digital Sky Survey.  The most accurate estimates of cosmological
parameters are obtained using a combined analysis of the 2dFGRS, SDSS
and the WMAP data \citep{Spergel:2003, Tegmark:2004, Sanchez:2006}.
According to these studies the matter density parameter is $\Omega_m =
0.27 \pm 0.02$, not far from the value $\Omega_m = 0.3$, suggested by
\citet{Ostriker:1995} as a concordant model.  The combined method
yields for the Hubble constant a value $h = 0.71 \pm 0.02$ independent
of other direct methods.  From the same dataset authors get for the
density of baryonic matter, $\Omega_b = 0.041 \pm 0.002$.  Comparing
both density estimates we get for the dark matter density $\Omega_{DM}
= \Omega_m - \Omega_b = 0.23$, and the dark energy density
$\Omega_\Lambda = 0.73$.  These parameters imply that the age of the
Universe is $13.7 \pm 0.2$ gigayears.

\subsection{The role of dark energy in the evolution of the 
Universe}

Studies of the Hubble flow in nearby space, using observations of type
Ia supernovae with the Hubble Space Telescope (HST), were carried out
by several groups.  The major goal of the study was to determine the
value of the Hubble constant.  As a by-product also the smoothness of
the Hubble flow was investigated.  In this project supernovae were
found up to the redshift (expansion speed) 20~000~km~s$^{-1}$.  This
project \citep{Sandage:2006} confirmed earlier results that the Hubble
flow is very quiet over a range of scales from our Local Supercluster
to the most distant objects observed. This smoothness in spite of the
inhomogeneous local mass distribution requires a special agent. Dark
energy as the solution has been proposed by several authors
(\citet{Chernin:2001, Baryshev:2001} and others).  Sandage emphasizes
that no viable alternative to dark energy is known at present, thus
the quietness of the Hubble flow gives strong support for the
existence of dark energy.

The vacuum dark energy has two important properties: its density
$\rho_v$ is constant, i.e. the density does not depend not on time nor
on location; and it acts as a repulsive force or antigravity (for
detailed discussions see \citet{Chernin:2003a}, \citet{Chernin:2006},
\citet{Chernin:2008}).

The first property means that in an expanding universe in the earlier
epoch the density of matter (ordinary + dark matter) exceeded the
density of dark energy. As the universe expands the mean density of
matter decreases and at a certain epoch the matter density and the
absolute value of the dark energy effective gravitating density were
equal. This happened at an epoch which corresponds to redshift $z
\approx 0.7$.  Before this epoch the gravity of matter decelerated the
expansion, after this epoch the antigravity of the dark energy
accelerated the expansion.  This is a global phenomenon - it happened
for the whole universe at once.

The density of the dark energy determines the expansion speed of the universe,
expressed through the (vacuum energy defined) Hubble constant
\begin{equation}
h_v = H_v/100 = \left({8\pi G \over 3}\rho_v\right)^{1/2} \approx 0.62.
\end{equation}
This value is rather close to the actually observed value of the Hubble
constant, given above, due to the dominance of the dark energy in the present epoch. 

The dark energy influences also the local dynamics of astronomical
bodies. Consider a virialised system as a group or cluster of galaxies. In the
first approximation the dynamics of the system can be treated as a point of
mass $M_m$ (index $m$ for matter).  The total force to a test particle moving
at a distance $R$ from the cluster center can be expressed as follows:
\begin{equation}
F(R) = -{G M_m \over R^2} +  {8\pi G \over 3} \rho_v R.
\end{equation}
The first term is due to the gravity of the cluster, the second term
is due to the antigravity of the dark energy in a sphere of radius $R$
around the cluster.  The antigravity corresponds to the effective mass
of the dark energy contained in the spherical volume of radius $R$:
$M_v = - {8\pi \over 3} \rho_v R^3$.  Near the cluster the gravity is
larger and determines the movement of test bodies. At large distance
the antigravity is larger, here stable orbits around the cluster are
impossible.  Both forces are equal at the distance $R_v = \left({3 M_m
    \over 8\pi \rho_v}\right)^{1/3}$; this distance can be called the
zero gravity distance \citep{Chernin:2008}.

The local effect of the dark energy to the dynamics of bodies has been
studied by Karachentsev, Chernin, Tully and collaborators. Using the
Hubble Space Telescope and large ground-based telescopes Karachentsev
determined accurate distances and redshifts of satellite galaxies in
the Local group and several nearby groups of galaxies
(\citet{Karachentsev:2002b}, \citet{Karachentsev:2003a},
\citet{Karachentsev:2003c}, \citet{Karachentsev:2006a},
\citet{Karachentsev:2007}, \citet{Karachentsev:2009},
\citet{Tully:2008}).  This study shows that near the group center up
to distance $R= 1.25$~\Mpc\ satellite galaxies have both positive and
negative velocities in respect to the group center, at larger distance
all relative velocities are positive and follow the Hubble flow.  The
distance 1.25~\Mpc\ corresponds exactly to the expected zero gravity
distance for groups of mass about $4\times 10^{12}$ solar masses. The
new total mass estimates are 3–-5 times lower than old virial mass
estimates of these groups, which leads to low density of matter
associated with these galaxies, $\Omega_m \approx 0.04$
(\citet{Karachentsev:2005a}).  If confirmed, this result may indicate
the presence of two types of dark matter: the matter associated with
galaxies, and a more smoothly distributed dark matter.

This test demonstrates that the dark energy influences both the local
and the global dynamics of astronomical systems.  For rich clusters of
galaxies the zero gravity distance is about 10~\Mpc, for rich
supercluster several tens \Mpc, which corresponds to the radius of
cores of rich superclusters. The antigravity of the dark energy
explains also the absence of extremely large superclusters: even the
richest superclusters have characteristic radii of about 50~\Mpc.

\subsection{Searches of Dark Matter particles}

{ In late 1970s and early 1980s it was clear that dark matter must
  be non-baryonic. The first natural candidate for DM particles was
  massive neutrino. Using astronomical constraints \citet{Szalay:1976}
  found upper mass limit of neutrinos as DM particles, $m_\nu <
  15$~eV. However, as discussed above, massive neutrinos (Hot Dark
  Matter) cannot form the dominating population of DM particles, since
  the large-scale-structure of the cosmic web would be in this case
  completely different from the observed structure.  For this reason
  hypothetical weakly interacting massive particles were suggested,
  which form the Cold Dark Matter. CDM model satisfies most known
  astronomical restrictions for the DM, as shown by
  \citet{Blumenthal:1984} and many others.

  Until recently it was thought that DM particles form a fully
  collisionless medium.  The ringlike DM distribution in the merging
  cluster Cl 0024+17 suggests that DM ``gas'' may have some
  collisional properties. \citet{Jee:2007wd} suggest that this cluster
  may serve as a very useful laboratory to address outstanding
  questions in DM physics.

  This review has focused only on gravitational aspects of
  DM. However, it is natural to assume that in the arguably most
  realistic cases, where the DM is provided by some sort of an
  elementary particle (beyond the Standard Model of particle physics),
  those particles have other than only gravitational couplings to the
  rest of the matter. If this is the case, the phenomenology of DM
  could in principle be much richer. Indeed, there has been a lot of
  recent activity in trying to detect DM particles in high precision
  nuclear recoil experiments. DAMA/LIBRA collaboration has announced a
  possible $\sim 9 \sigma$ detection of the DM induced signal visible
  as an expected annual modulation in the observational data,
  \citet{Bernabei:2010tw}.

  Also, a multitude of astrophysical observations have been used to
  search for the indirect hints for the existence of the DM
  particles. Particularly interesting is the cosmic ray positron
  anomaly as revealed by the measurements of the PAMELA
  \citep{Adriani:2009ye}, Fermi \citep{Abdo:2009qf}, and HESS
  \citep{Aharonian:2009fu} experiments. This anomaly could be an
  indirect indication for the existence of the annihilating or
  decaying DM particle with a mass at the TeV scale.

Results from the neutrino oscillation experiments require at least one
of the neutrinos to have a mass not less than $\sim 0.05$ eV
(e.g. \citet{Dolgov:2002fk}).  This immediately implies that the corresponding
density parameter $\Omega_{\nu}\gtrsim 0.001$, i.e. approaching the
density parameter of the baryons visible in the form of stars!
Although neutrinos cannot form the dominant component of the DM, due
to reasons discussed above, it shows that the general idea of the
existence of DM in Nature is surely not a fiction.

  For obvious reasons it is not possible here to give a full account
  of all the activities of DM searches that belong to the rapidly
  developing field of astroparticle physics. Instead we advise the
  reader to consult the excellent recent review papers by
  \citet{Bertone:2005bv}, \citet{Bergstrom:2009kb},
  \citet{Feng:2010jw}, and \citet{Profumo:2010qo}.}

\section{Conclusions}

The discoveries of dark matter and the cosmic web are two stages of a
typical scientific revolution \citep{Kuhn:1970, Tremaine:1987}.  As
often in a paradigm shift, there was no single discovery, new concepts
were developed step-by-step.

First of all, actually there are two dark matter problems -- the local
dark matter close to the plane of our Galaxy, and the global dark
matter surrounding galaxies and clusters of galaxies.  Dark matter in
the Galactic disk is baryonic (faint stars or jupiters), since a
collection of matter close to the galactic plane is possible, if it
has formed by contraction of pre-stellar matter towards the plane and
dissipation of the extra energy, that has conserved the flat shape of
the population. The amount of local dark matter is low; it depends on
the mass boundary between luminous stars and faint invisible stars.

The global dark matter is the dominating mass component in the
Universe; it is concentrated in galaxies, clusters and superclusters,
and populates also cosmic voids. Global dark matter must be
non-baryonic, its density fluctuations start to grow much earlier than
perturbations in the baryonic matter, and have at the recombination
epoch the amplitude large enough to form all structures seen in the
Universe.  Initially neutrinos were suggested as particles of dark
matter (hot dark matter), but presently some other weakly interacting
massive particles, called cold dark matter, are preferred.

Recently direct observational evidence was found for the presence of
Dark (or vacuum) Energy.  New data suggest that the total
matter/energy density of the Universe is equal to the critical
cosmological density, the density of baryonic matter is about 0.04 of
the critical density, the density of dark matter is about 0.23 of the
critical density, and the rest is dark energy.

A number of current and future astronomical experiments have the aim
to get additional data on the structure and evolution of the Universe
and the nature and properties of dark matter and dark energy.  Two
astronomical space observatories were launched in 2009: the Planck CMB
mission and the Herschel 3.5-m infrared telescope.  The main goal of
the Planck mission is to measure the CMB radiation with a precision
and sensitivity about ten times higher than those of the WMAP
satellite.  This allows us to estimate the values of the cosmological
parameters with a very high accuracy.  The Herschel telescope covers
the spectral range from the far-infrared to sub-millimeter wavelengths
and allows us to study very distant redshifted objects, i.e young
galaxies and clusters \citep{Cooray:2010qa}.

Very distant galaxies are the target of the joint project GOODS -- The
Great Observatories Origins Deep Survey.  Observations are made at
different wavelengths with various telescopes: the Hubble Space
Telescope, the Chandra X-ray telescope, the Spitzer infrared space
telescope, and by great ground-based telescopes (the 10-m Keck
telescope in Hawaii, the 8.2-m ESO VLT-telescopes in Chile). Distant
cluster survey is in progress in ESO \citep{White:2005}.

NASA and U.S. Department of Energy formed the Joint Dark Energy
Mission (JDEM) and proposed a space observatory SNAP (the SuperNova
Acceleration Probe) to detect and obtain precision photometry,
light-curves and redshifts of over 2000 type Ia supernovae over the
redshift range $0.5 < z < 1.7$ to constrain the nature of dark energy.

The largest so far planned space telescope is The James Webb Space
Telescope (JWSP) -- a 6.5-m infrared optimized telescope, scheduled
for launch in 2013.  The main goal is to observe first galaxies that
formed in the early Universe.

To investigate the detailed structure of our own Galaxy the space
mission GAIA will be launched in 2011.  It will measure positions,
proper motions, distances and photometric data for 1 billion stars,
repeatedly.  Its main goal is to clarify the origin and evolution of
our Galaxy and to probe the distribution of dark matter within the
Galaxy.

The story of the dark matter and dark energy is not over yet -- we
still do not know of what non-baryonic particles the dark matter is
made of, and the nature of dark energy is also unknown. We even do not
know is a radical change in our understanding of the Newton and
Einstein theories of gravitation needed. All these problems are
challenges for physics.  So far the direct information of both dark
components of the Universe comes solely from astronomical
observations.

\subsection*{Acknowledgments}

The study of dark matter and large-scale structure of the Universe in
Tartu Observatory is supported by Estonian Science Foundation grant
6104, and Estonian Ministry for Education and Science grant TO
0060058S98. The author thanks Astrophysikalisches Institut Potsdam
(using DFG-grant MU 1020/11-1), ICRANet and the Aspen Center for
Physics for hospitality, where part of this study was performed, and
Elmo Tempel and Triin Einasto for help in preparing the bibliography.
Fruitful discussions with Arthur Chernin, Maret Einasto, Gert H\"utsi,
Enn Saar, Alar Toomre, Virginia Trimble, Sidney van den Bergh, and
the editor of the series Bozena Czerny helped to improve the quality
of the review.

\subsection*{Glossary}

{\bf Acoustic peaks}: features in the angular spectrum of the CMB
radiation (and corresponding peak in the correlation function of
galaxies) due to the acoustic oscillations of the hot gas in the young
universe before the recombination (sound waves driven by cosmic
perturbations of the hot plasma).

{\bf Baryonic dark matter}: dark matter composed of baryons - protons,
neutrons and bodies made of baryons. Candidates for baryonic dark
matter are MACHOs, brown dwarfs, Jupiters and non-luminous gas.

{\bf Biased galaxy formation}: a model of galaxy formation suggesting
that galaxies form only in high-density regions, whereas in
low-density regions the matter remains in gaseous (pre-galactic) form.

{\bf Big Bang}: a model of the formation of the Universe from an
extremely dense and hot state with followed by very rapid expansion
(inflation) and more moderate expansion today.

{\bf Bulge}:  a spheroidal population of galaxies consisting of old
medium metal-rich stars.

{\bf Clusters of galaxies}: largest gravitationally bound aggregates of galaxies,
containing typically more than 30 galaxies and having diameters up to 5
megaparsecs. They consist of visible galaxies, hot intracluster medium and
invisible dark matter

{\bf CMB fluctuations}: fluctuations of the temperature of the CMB
radiation, seen on the sky as small anisotropies - deviations from the
mean temperature. Temperature fluctuations are related to fluctuations
of the density of matter, from overdense regions of the primordial
plasma due to gravitational clustering all large astronomical objects
formed, such as galaxies, clusters of galaxies.

{\bf CMB radiation - Cosmic microwave background radiation}:
electromagnetic radiation filling the universe, and formed during the
hot stage of the evolution of the universe just before recombination
of light chemical elements which makes the universe transparent for
radiation. As the universe expanded both the plasma and radiation
cooled, so the temperature of the CMB radiation is presently about 2.7
K, seen in the microwave range of the spectrum.

{\bf Cold dark matter (CDM)}: a variety of dark matter which consists
of particles having small velocities relative to the speed of light,
which allows these particles to form density enhancements called dark
halos. Candidate CDM particles, such  as axions, or
supersymmetric particles are proposed, none of those experimentally detected yet.

{\bf Corona}: an extended and nearly spherical population surrounding
galaxies and clusters of galaxies, consisting of hot X-ray emitting
gas and dark matter.

{\bf Cosmic inflation}: theoretically predicted period in the
evolution of the Universe during which the whole Universe expanded
exponentially after the Big Bang, driven by the negative-pressure
(antigravity) of the vacuum energy.

{\bf Cosmic web}: a spider-web-like distribution of galaxies in the
universe along filaments and superclusters, also called
supercluster-void network.

{\bf  Critical density of the universe}: the density of
  matter/energy in the universe that separates the two spatial
  geometries - opened and closed; with the dividing line providing
  flat geometry.

{\bf Dark ages}: the period in time evolution of the Universe after the
recombination and before the formation of first stars and galaxies.

{\bf Dark halo (often shortly halo)}: an extended and nearly spherical
population surrounding galaxies and clusters of galaxies, consisting
of dark matter.

{\bf Dark matter}: hypothetical matter, whose presence can be detected
from its gravity only.

{\bf Disk}: a relatively flat population in galaxies, consisting of
stars of various age and metallicity.

{\bf Elliptical galaxies}: galaxies consisting mainly of old
populations, having an approximately ellipsoidal shape (bulge, halo).

{\bf Filament}: a chain of galaxies and galaxy systems (groups or clusters). 

{\bf Galactic evolution}: the change of physical parameters of
galaxies and their populations in time. As parameters such quantities
are considered as age, stellar content, chemical composition,
photometric properties (color), mass-to-luminosity ratios etc.

{\bf Galactic models}: mathematical models which describe quantitatively the
structure of galaxies and their populations - spatial density, rotation,
physical properties (colors, luminosities, mean ages of stars etc).

{\bf Galactic populations}: populations of stars or gas of similar
age, composition (metallicity), spatial distribution and kinematical
properties (velocity dispersion of stars, rotation velocity around the
galactic center). Main galactic populations are nucleus, disk, young
population, bulge, stellar halo, dark halo (or corona).

{\bf Galaxies}: large systems of stars and interstellar matter,
consisting of one or more galactic populations.

{\bf Global dark matter}: dark matter surrounding galaxies and
clusters of galaxies and populating voids between galaxy
systems. Most, if not all, global dark matter is non-baryonic.

{\bf Gravitational clustering}: the clustering of matter due to
gravity - matter concentrates toward overdense regions and flows away
from underdense regions in space.

{\bf Gravitational lensing}: the bending of the light from a distant
source around a massive object like a cluster of galaxies.

{\bf Gravitational microlensing}: gravitational lensing where the
amount of light received from a background object changes in
time. Microlensing allows us to study low-mass objects (brown and red
dwarfs, planets) that emit little or no light.

{\bf Groups of galaxies}: smallest gravitationally bound systems of
galaxies, containing typically fewer than 30 galaxies and having
diameters of 1 to 2 megaparsecs.

{\bf Hot dark matter (HDM)}: a variety of dark matter which consists
of fast moving particles, close to the speed of light. Neutrinos are
the only experimentally confirmed HDM particles.

{\bf Hubble's law}: the equation stating that the recession velocity
of galaxies from the earth is proportional to their distance from us;
the constant of proportionality is called the Hubble constant.  Recent
data suggest a value of the Hubble constant $H_0 = 71$ (km/s)/Mpc.

{\bf Initial mass function (IMF)}: an empirical function that
describes the mass distribution of a population of stars at the epoch
of star formation.

{\bf Jupiter}: in cosmology a giant planet-type body with mass less
than a critical value needed to start nuclear reactions inside the
body.

{\bf Lambda CDM model}: (Lambda-Cold Dark Matter) model or the
concordant model is a model of the universe which has critical
cosmological density and contains three main components of the
matter/energy: baryonic matter, dark matter and dark energy; the last
component is also called the cosmological constant Lambda.

{\bf Large scale structure of the Universe}: the characterization of the
distribution of large-scale structures (galaxies and systems of galaxies) in
space. Galaxies are concentrated to filaments forming superclusters, the space
between filaments contains no galaxies (cosmic voids).

{\bf Local dark matter}: dark matter in the Solar vicinity near the
plane of the Galaxy. Local dark matter is baryonic since collisionless
non-baryonic dark matter cannot form a flattened disk.

{\bf MACHO - massive compact halo object}: astronomical body that
emits little or no radiation and can be detected by its gravitation
only. MACHO's were suggested as an alternative to non-baryonic dark
matter candidate in galactic dark halos.

{\bf Megaparsec (Mpc)}: a measurement of distance in the Universe, equal to one
million parsecs or 3.26 million light years ($3.086 \times 10^{22}$ meters).

{\bf MOND - modified Newtonian dynamics}: a theory that proposes a
modification of the Newton's Law of Gravity to explain flat rotation
curves of galaxies.

{\bf Morphology of galaxies}: structural properties of galaxies, like
the presence of different stellar or gaseous populations, their color,
metal content, age, as well as size and shape, kinematical properties
etc.  Morphological properties of galaxies are explained by formation
and evolution of their populations.

{\bf N-body simulation}: simulation of the formation and evolution of
structures of the universe using massive particles under the influence
of gravity, sometimes also other forces to take into account the
evolution of the gas to stars.

{\bf Non-baryonic dark matter}: dark matter composed of non-baryonic
particles, i.e. it has no atoms and bodies composed of atoms, such as
various chemical elements.

{\bf Primordial nucleosynthesis}: the process of forming nuclei of
light elements from protons and neutrons. This process happens in a
short time interval when the temperature of the universe and its
density are in limits which allows this process to occur (between 3
and 20 minutes after the Big Bang).

{\bf Paradigm}: the set of ideas and practices that define a
scientific discipline during a particular period of time.

{\bf Recombination}: the period in the evolution of the universe when
temperature was low enough to allow nuclei of light elements
(hydrogen, helium, lithium) bound electrons around the nucleus and to
form neutral atoms of these elements.

{\bf Scientific revolution}: a period in the history of science when
new observations or experiments led to a rejection of older doctrines
(paradigms) and formation of new ideas (paradigms), also called
paradigm shift.

{\bf Spiral galaxies}: galaxies containing a flat disk of young stars and
interstellar matter, which form spiral arms.

{\bf Standard cosmological model (LCDM model)}: currently accepted
cosmological model with parameters: Hubble constant 71 km/s per Mpc,
baryonic matter density 0.04, dark matter density 0.23, and dark
energy (or cosmological constant) density 0.73, all in units of the
critical cosmological density.

{\bf Stellar halo}: a population of old metal-poor stars in galaxies.

{\bf Structure formation}: the formation of the structure of the
universe in the process of gravitational clustering which attracts
matter to higher density regions (superclusters) and makes voids
emptier.

{\bf Supercluster}: density enhancement in the cosmic web, consisting
of one or more clusters of galaxies and galaxy/cluster filaments with
voids between them. Superclusters can be defined as the largest
non-percolating systems of galaxies and clusters/groups in the
Universe.

{\bf Supernova}: an explosion of a star  which causes a burst of radiation
that for some time (from weeks to months) outshines an entire galaxy. 

{\bf Universe}: everything that physically exists - all forms of matter, energy,
space and time, governed by physical laws and constants. An alternative
definition is ``our Universe'', which means the Universe we live in and
postulates the presence of many disconnected ``universes'' as the multiverse.  

{\bf Virial theorem}: a theorem in mechanics that provides an equation
that relates the average over time of the total kinetic energy with
that of the total potential energy.

{\bf Void}: a region between galaxies and systems of galaxies devoid of visible
objects.

{\bf WIMP - weakly interacting massive particles}: hypothetical
particles, serving as candidates to non-baryonic dark matter. They
interact through gravity and possible through nuclear force no
stronger than the weak force. Thus their interaction with
electromagnetic radiation is so weak that their density evolution can
start much earlier than for baryonic particles.

{\small

}

\end{twocolumn}

\end{document}